\documentclass[journal]{IEEEtran}
\AtBeginDocument{%
  }

\usepackage{amsmath}

\usepackage{url}
\usepackage{amssymb}
\usepackage{amsfonts}

\usepackage{colortbl}
\usepackage{hhline}
\usepackage{relsize}
\usepackage{enumitem}
\usepackage{comment}

\usepackage[table,xcdraw,dvipsnames,svgnames,x11names]{xcolor}

\usepackage[hidelinks,colorlinks=true,linkcolor=Emerald,citecolor=cyan,urlcolor=black]{hyperref}

\newcommand{\ignore}[1]{ }
\usepackage{graphicx}
\usepackage{textcomp}

\usepackage{float}

\usepackage{colortbl}
\definecolor{best}{RGB}{220,240,255}
\definecolor{second}{RGB}{235,235,235}
\usepackage{pifont}
\usepackage{marvosym}
\usepackage{ragged2e}

\usepackage{tikz}

\usepackage{float}
\usetikzlibrary{positioning,arrows.meta}

\usepackage{pgfplots}
\pgfplotsset{compat=1.18}
\usetikzlibrary{plotmarks}
\definecolor{cAPUF4}{rgb}{0.8500, 0.3250, 0.0980}
\definecolor{cAPUF5}{rgb}{0.0000, 0.4470, 0.7410}
\definecolor{cAPUF6}{rgb}{0.7000, 0.4470, 0.7410}
\definecolor{cAPUF7}{rgb}{0.9000, 0.7000, 0.5410}
\definecolor{cAPUF8}{rgb}{0.1000, 0.7000, 0.4100}
\definecolor{cAPUF16}{rgb}{0.3500, 0.3900, 0.1600}
\definecolor{cLFSR}{rgb}{0.4660, 0.6740, 0.1880}
\definecolor{cSobolC}{rgb}{0.4940, 0.1840, 0.5560}
\definecolor{cSobolF}{rgb}{0.2000, 0.5600, 0.7800}
\definecolor{cTRUE}{rgb}{0.8800, 0.3200, 0.1700}

\usepackage{multirow}
\usepackage{vcell}

\usepackage{courier}
\usepackage{balance}

\usepackage{hhline}
\usepackage{arydshln}

\usepackage{textcomp}

\usepackage{lipsum}
\usepackage{setspace}
\usepackage{booktabs}

\usepackage{graphicx}

\usepackage{algorithm}
\usepackage{algpseudocode}
\usepackage{tikz,xcolor}

\definecolor{lime}{HTML}{A6CE39}
\DeclareRobustCommand{\orcidicon}{%
	\begin{tikzpicture}
	\draw[lime, fill=lime] (0,0) 
	circle [radius=0.14] 
	node[white] {{\fontfamily{qag}\selectfont \tiny ID}};
	\draw[white, fill=white] (-0.0625,0.095) 
	circle [radius=0.007];
	\end{tikzpicture}
	\hspace{-2mm}
}

\foreach \x in {A, ..., Z}{%
	\expandafter\xdef\csname orcid\x\endcsname{\noexpand\href{https://orcid.org/\csname orcidauthor\x\endcsname}{\noexpand\orcidicon}}
}

\usepackage{soul}
\sethlcolor{Azure2} %Azure2 LightGray

\newcommand{\Design}{\textbf{\texttt{APUF-SNG}}}

\begin{document}
%\sloppy

\title{ 
%\Mehcolor{\Design: Turning Arbiter-PUF Metastability into Controllable Stochastic Bit-Streams}
\Design: A Low-Cost Arbiter-PUF Stochastic Number Generator with Tunable Output Probability
%\Design: A Low-Cost Arbiter-PUF Architecture for\\ True-Random Stochastic Bit-Stream Generation
%\vspace{-.5em}
}

\author{Shelby Williams\orcidS{}, Mehran~Moghadam\orcidA{}, M.~Hassan~Najafi\orcidD{},\\ Sercan~Aygun\orcidC{}, Arman Roohi\orcidE{}, and Gourav Datta\orcidF{}
\thanks{This work is supported by NSF grants 2339701, 2609436, 2616837, NASA grant 80NSSC25C0335, Vernon $\&$ Ruby Langlinais Non-Endowed Research Fund, Lockheed Martin Corporation Endowed Professorship Fund, and gifts from NVIDIA and Google. Mehran Moghadam and M. Hassan Najafi are with the ECSE %Electrical, Computer, and Systems Engineering 
Department, Case Western Reserve University, Cleveland, OH, USA. E-mail:\{moghadam, najafi\}@case.edu. Sercan Aygun is with the School of Computing and Informatics, University of Louisiana at Lafayette, Lafayette, LA, USA. E-mail:\{c00591145, sercan.aygun\}@louisiana.edu. Arman Roohi is with the ECE Department at the University of Illinois Chicago, Chicago, IL, USA. E-mail: aroohi@uic.edu. Gourav Datta is with the ECSE Department, Case Western Reserve University, Cleveland, OH, USA. E-mail: gourav.datta@case.edu.} % <-this % stops a space
%\thanks{\scriptsize A preliminary version of this work appeared as~\cite{ID_VSA_ISLPED25}.}
%\vspace{-5pt}
\vspace{-1.em}
}

%\markboth{IEEE TRANSACTIONS ON VERY LARGE SCALE INTEGRATION (VLSI) SYSTEMS, Vol.~XX, No.~X, XXXX~2026}%

\maketitle

\begin{abstract}
Stochastic computing (SC) encodes numerical values as probabilistic bit-streams and enables arithmetic operations using simple logic gates, offering a promising paradigm for low-cost, fault-tolerant, and highly parallel computation. However, the accuracy and efficiency of SC systems depend on the quality of the stochastic number generator (SNG). Conventional SNGs rely on pseudo-random or quasi-random generators, whose deterministic nature introduces periodicity, correlation, and bias, limiting achievable accuracy and robustness. %in sequential computations.
This paper proposes a low-cost arbiter physical unclonable function (APUF)-based SNG that operates in near-metastable delay conditions to generate true-random stochastic bit-streams with controllable output probabilities. By selecting input challenges that drive the path delay difference toward a target value, the resulting output probability is governed by intrinsic device noise, eliminating the need for conventional random-number generators and comparator-based SNGs. The proposed design inherits
the low hardware overhead and statistical independence properties
of the APUF structure. 
A probabilistic model and a challenge selection method
%Monte Carlo-based challenge selection methodology 
are developed to characterize and control the stochastic behavior.
An 8-stage \Design{} matches the 1\% mean absolute error of state-of-the-art true-random SNGs at 1024-bit streams and occupies 116$\mu m^2$ in 45$nm$, 2.5$\times$ and 7$\times$ smaller than the TRNG-based and Sobol-based generator designs, respectively.
%Experimental results demonstrate competitive accuracy and improved energy efficiency compared to state-of-the-art pseudo-random, quasi-random, and true random bit-stream generators. 
%Experimental results demonstrate that the proposed APUF-SNG achieves a mean absolute error of 1.01\% at a bit-stream length of 1024, reduces hardware area by approximately 72\% compared to Sobol-based generators, and improves energy efficiency by over 1000$\times$ relative to state-of-the-art TRNG implementations.
%The results establish the proposed APUF-SNG as a compact, energy-efficient, and high-quality solution for true-random stochastic bit-stream generation, enabling scalable and reliable SC systems.
\end{abstract}

\begin{IEEEkeywords}
Physical unclonable function, true-randomness, stochastic computing, bit-stream generation, edge AI.
\end{IEEEkeywords}

%\vspace{-0.5em}
\section{Introduction}
%\vspace{-0.25em}

\IEEEPARstart{S}{tochastic} computing (SC) is a re-emerging computing paradigm that encodes data as random bit-streams, where the value of a number is represented by the probability of observing a logic `1'~\cite{Weikang_2011,alaghi2017promise}. Unlike traditional binary computing, which relies on positional number representations requiring complex multi-bit arithmetic units, SC performs computation using remarkably simple logic gates: multiplication of two values reduces to a bitwise \texttt{AND} operation, scaled addition is implemented by a multiplexer (\texttt{MUX}), and other complex functions use comparably minimal hardware~\cite{TranSC_Moghadam_TVLSI2026}. These properties make SC particularly attractive for resource-constrained and fault-tolerant applications, including image and video processing~\cite{11261774,p2lsg-asp-dac}, machine learning inference~\cite{9159931}, signal processing~\cite{10153851}, and neural network acceleration~\cite{9116390}. %where the inherent noise tolerance of SC and its low hardware cost provide tangible advantages over conventional binary designs.

A critical %and often underappreciated
bottleneck in SC systems is the stochastic number generator (SNG), the unit that converts binary-encoded data into random bit-streams. The SNG typically consists of two components: a random number generator (RNG) and a binary comparator. At each clock cycle, the comparator produces a single output bit by checking whether the target input value exceeds the current random number provided by the RNG.   
The overall quality of the generated bit-streams, the uniformity of the distribution of `1's and `0's and the degree of correlation between generated streams, directly govern the accuracy, latency, and energy efficiency of the computation. 
A poor SNG lengthens the bit-streams needed to converge, which raises latency and energy.
%A poor-quality SNG introduces what is known as \textit{random fluctuation} error, causing the SC system to require longer bit-streams to converge to an accurate result, which increases both latency and energy consumption.

\begin{figure}[!t]
%\vspace{0.75em}
\centering
\includegraphics[width=\columnwidth]{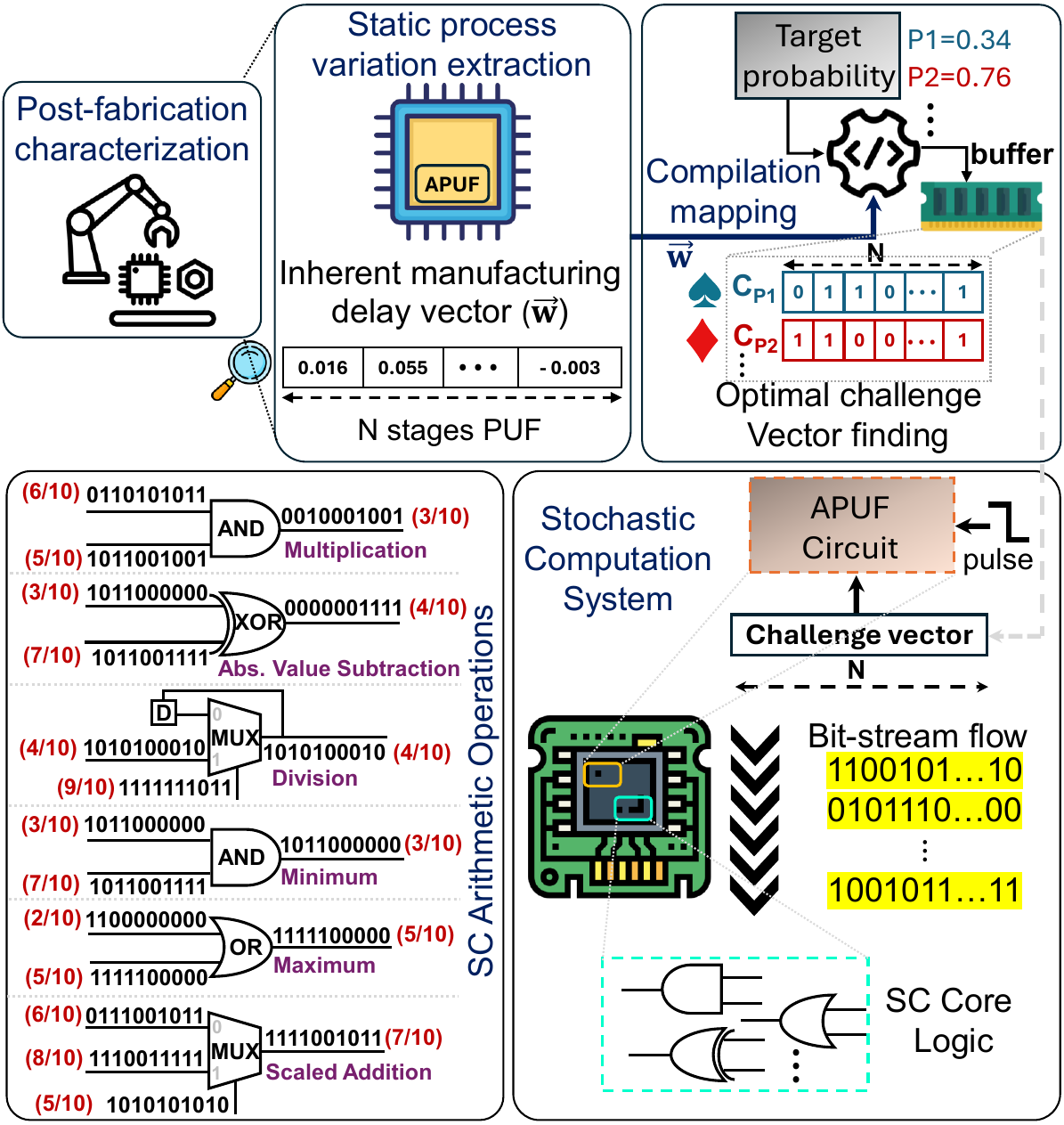}
\vspace{-2em}
\caption{An overview of the proposed \Design~design. %[\red{Cross-reference needed}]
}
\label{fig:APUF_SNG}
\vspace{-1.0em}
\end{figure}

%Historically, 
%The dominant approach for random number generation in SC has relied on 
Most SC designs generate random numbers with
pseudo-random number generators (PRNGs), commonly implemented as linear-feedback shift registers (LFSRs)~\cite{Weikang_2011}. LFSRs are compact and easy to implement in hardware, but their deterministic and periodic nature introduces statistical biases and inter-stream correlations that degrade computation accuracy~\cite{alaghi2017promise}. %To address these limitations, 
State-of-the-art (SOTA) works replaced LFSRs with quasi-random number generators (QRNGs) based on low-discrepancy (LD) sequences, such as Sobol~\cite{liu2017energy}, Halton~\cite{Alaghi_Fast}, and van der Corput~\cite{p2lsg-asp-dac} sequences. LD sequences distribute their values more uniformly over the sample space, allowing SC bit-streams to converge to the target probability more quickly, reducing the required bit-stream length and thus improving energy efficiency~\cite{liu2017energy}.
%Despite these advances, 
Both PRNG- and QRNG-based SNGs %share a fundamental limitation: they 
are \textit{deterministic}: their sequences %they produce 
are \textit{predictable} from seed, %from their initial state, 
which introduces biases and periodic correlations that place a hard ceiling on achievable accuracy.
Furthermore, the high hardware cost of these generators %, which are built from comparators, direction vector arrays, prime-base counters, and associated combinational logic, 
can offset the hardware simplicity that motivates the use of SC. % in the first place.

True-random number generators (TRNGs) offer a different foundation %a principally different and more robust foundation 
for SC bit-stream generation. Rather than relying on deterministic algorithms, TRNGs derive randomness from inherently unpredictable physical processes such as thermal noise, electronic jitter, and quantum phenomena. This intrinsic non-determinism eliminates periodic correlations, minimizes inter-stream bias, and provides high entropy, all of which are highly desirable properties for SC. Most recent works exploring TRNGs for SC include designs based on resistive RAM (RRAM) devices~\cite{SC_IMC_DAC2025} and ring-oscillator-based true-random bit-stream generation~\cite{TRUE_BSG_ISCAS2025}. %generators that can produce random bits at high speeds~\cite{10364295}. 
These efforts demonstrate that TRNG-based SC bit-stream generators can achieve accuracy comparable % deterministic bit-stream generators while reducing energy consumption by nearly an order of magnitude compared 
to LFSR- and Sobol-based designs. However, TRNG-based SNG designs %remain rare, and none combines true randomness, low cost, and controllable output probability.
remain relatively unexplored, and 
implementations that simultaneously offer true randomness, low hardware overhead, and controllable output probability remain an open and important research problem.

Physical unclonable functions (PUFs) are a class of hardware security primitives that exploit the intrinsic and uncontrollable manufacturing variations in integrated circuits to generate device-unique responses~\cite{11460260}. Among the many PUF architectures, the \textit{arbiter PUF (APUF)} is particularly well-suited to silicon implementation due to its compact, regular structure~\cite{11326004}. In a standard APUF, two electrically equivalent signal paths race through a cascade of challenge-controlled \texttt{MUX} stages, and a latch at the output determines which path arrives first. Because fabrication-induced delay variations are random and practically impossible to predict or replicate, the challenge-response behavior of an APUF forms a device-specific fingerprint. Conventional APUF design %deliberately 
avoids operating conditions where the two path delays are nearly equal, since such near-balanced conditions introduce metastability and unpredictable output behavior. Prior works demonstrated, however, that these metastable challenges (for which the path delay difference approaches zero) can serve as a source of true randomness, with each re-evaluation of a fixed near-metastable challenge producing an independent, nearly unbiased random bit~\cite{10150018}. %This insight opens a fundamentally new perspective on APUFs: rather than treating near-balanced delay conditions as a design hazard to be avoided, they can be deliberately engineered and selected as a probabilistic computation resource.
These designs treat such instability as a security feature: DTA-PUF derives responses from timing errors in pipelined cores~\cite{tsiokanos2021dtapuf}, and dual-mode circuits switch between PUF and TRNG operation~\cite{11421416, 9935279}. None controls the output probability. We instead calibrate the near-balanced region so that a stored challenge produces any target output probability, which is what an SNG requires.

In this work, we propose, to the best of our knowledge, the first APUF-based true-random stochastic bit-stream generator for SC systems. Fig.~\ref{fig:APUF_SNG} provides an overview of the proposed %shows a high-level  architecture of the 
\Design{} design. %based framework for true random bit-stream generation. %, enabling stochastic-bitstream computing. 
Our approach reinterprets the APUF as a \textit{time-to-probability conversion primitive}: by %carefully 
selecting %input 
\textit{challenges} such that the cumulative path delay difference $\Delta$ takes a specific target value, we can control the probability that the arbiter resolves to `1' on any given evaluation. Specifically, under additive Gaussian noise arising from thermal noise, jitter, and device mismatch, the output probability follows $P(S=1) = \Phi(\frac{\Delta}{\sigma})$, where $\Phi(\cdot)$ is the standard normal cumulative distribution function (CDF) and $\sigma$ characterizes the noise magnitude. By mapping each %desired SC
input value to a challenge that produces the corresponding output probability, we generate stochastic bit-streams directly from the physical timing dynamics of the PUF, with no RNG or comparator. %without requiring any of the combinational logic overhead that burdens conventional SNG designs. 
The proposed architecture inherits several favorable properties of APUFs: \ding{172} any two bit-streams generated from distinct challenges are statistically uncorrelated, %($\text{SCC} \approx 0$), 
satisfying a key requirement for accurate SC multiplication and many other SC operations; \ding{173} the randomness is rooted in physical process variation and thermal noise rather than a deterministic algorithm; and \ding{174} the hardware footprint of the core generator is minimal, consisting only of the PUF delay chain and its sensing latch.

We present an analytical delay-and-noise model, together with a challenge selection method.
%a Monte Carlo methodology for challenge space exploration and optimal challenge selection. %Extensive 
Evaluations %are conducted to 
characterize the statistical quality of the generated bit-streams. The proposed generator is compared against SOTA stochastic bit-stream generation techniques, including LFSR-based PRNG, Sobol-based QRNG, and recent true-random approaches such as TRUE-BSG~\cite{TRUE_BSG_ISCAS2025} and IMSNG~\cite{SC_IMC_DAC2025}. % in terms of accuracy, hardware cost, and energy efficiency. %ring-oscillator-based TRNG bit-stream generator~\red{[X]}.
The key contributions of this work are summarized as follows:
\begin{itemize} [leftmargin=0pt, itemindent=1em, 
                labelsep=0.3em, labelindent=0pt,
                itemsep=4pt, label=\ding{226}]

\item \textbf{\textcolor{Navy}{APUF-Based Stochastic Bit-Stream Generation:}} 
We introduce \Design, %, for the first time, 
an APUF-based true-random stochastic bit-stream generator % (\Design) 
that directly converts device-level delay variations into probabilistic outputs. A formal probabilistic model is developed to map path delay differences to output probabilities through a noise-driven decision process, enabling continuous and fine-grained probability generation. %without conventional random-number generators.

\vspace{-0.35em}
\item \textbf{\textcolor{Navy}{Probability Control via Challenge Selection:}} 
We propose a challenge selection method that identifies near-metastable operating points corresponding to arbitrary target probabilities. This approach enables precise probability tuning and eliminates the need for comparator-based encoding and external RNGs employed in conventional SC systems.

%\item \textbf{Challenge Space Characterization:} We introduce a statistical challenge search methodology that identifies near metastable challenges corresponding to any target output probability, enabling the generation of stochastic bit-streams for arbitrary SC input values.

%We perform large-scale statistical characterization of the APUF challenge space across multiple stage configurations, identifying subsets of near-metastable and timing-efficient challenges. These subsets enable high-quality stochastic bit-stream generation with improved randomness, reduced correlation, and enhanced operating speed.

\vspace{-0.35em}
\item \textbf{\textcolor{Navy}{Hardware Cost and Accuracy Evaluation:}} 
We provide a comprehensive evaluation of the proposed generator against SOTA PRNG-, QRNG-, and TRNG-based approaches in terms of accuracy, area, power, and energy consumption, demonstrating competitive
%or superior 
efficiency at significantly lower hardware
cost. %The proposed design achieves a mean absolute error of 1.01\%, reduces hardware area by approximately 72\%, and improves energy efficiency by over 1000$\times$, demonstrating a superior trade-off between accuracy, randomness quality, and hardware efficiency.

%\item \textbf{Time-to-Probability Mapping:} We establish a formal probabilistic model that maps path delay difference to output probability via a Gaussian noise-driven decision process, enabling a continuous transition between deterministic and stochastic regimes.

%\item \textbf{Stochastic Arbiter-PUF Architecture:} We propose a 64-stage arbiter-PUF design operating in a near-metastable regime, capable of generating stochastic bitstreams with controllable output probabilities.

%\item \textbf{Latch-Level Physical Modeling:} We develop a StrongARM latch-based timing and noise model that captures regeneration dynamics and explains stochastic behavior as a function of input delay imbalance and noise.

%\item \textbf{Logic-Free Stochastic Multiplication:} We demonstrate that stochastic multiplication can be realized through timing-driven conditional evaluation, eliminating the need for conventional Boolean logic gates.

%\item \textbf{Challenge Space Optimization:} We introduce a statistical framework for identifying near-metastable and timing-efficient challenges using large-scale sampling and ranking.

%\item \textbf{Unified Security–Computation Paradigm:} We present a dual-use hardware primitive that simultaneously supports PUF-based identification and stochastic computation, opening new directions in hardware–algorithm co-design.
%\vspace{-0.5em}
\end{itemize}

\begin{figure*}[!t]
\centering
%\vspace{-0.75em}
%\includegraphics[width=1.0\linewidth, trim=555 50 220 360, clip]{Figures/Figure2.pdf}
\includegraphics[width=0.8\linewidth]{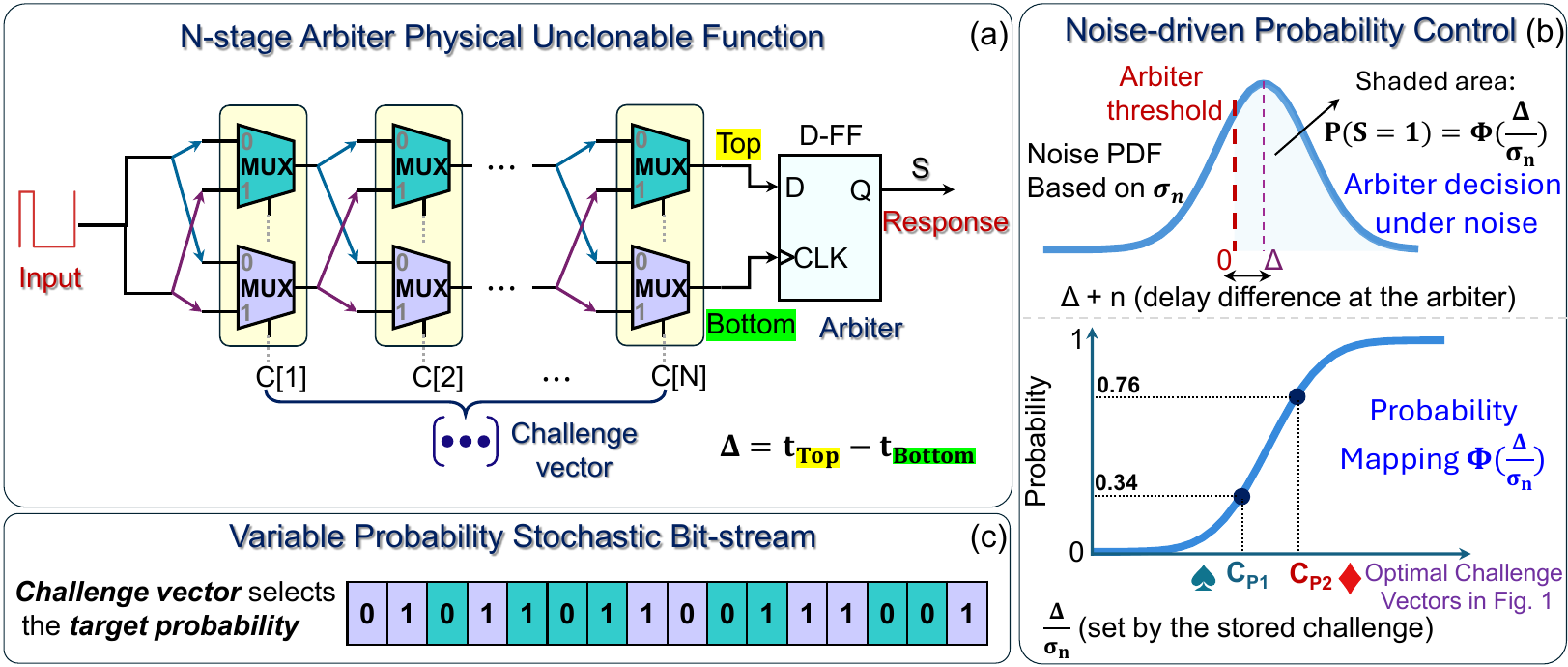}
\vspace{-0.5em}
\caption{
(a) An $N$-stage APUF structure, (b) arbiter decision under noise together with probability mapping, and (c) resulting SC bit-stream.
%(a) An N-Stage APUF as a True-Random Stochastic Bit-stream Generator, (b) the  Metastable Region of the APUF and (c) %An Example of a Stochastic Bitstream approximating $x=0.5$. 
%An Example of 
%A True-Random Bit-Stream Generated Near Metastable Boundary ($P(S=1) \approx 0.5$).
%probability value $x$=$9/16$
}
%\Description{An  Arbiter-PUF True-Random Stochastic Bitstream Generator}
\label{fig:figure2}
\vspace{-0.5em}
\end{figure*}

%\vspace{-0.5em}

%The remainder of this paper is organized as follows. Section~II reviews the fundamentals of APUFs and their stochastic properties. Section~III presents the proposed APUF-based stochastic bit-stream generator. Section~IV develops the probabilistic modeling framework for mapping delay variations to output probabilities and the associated challenge selection methodology. Section~V provides simulation results and a comprehensive performance evaluation, including accuracy, hardware cost, and energy efficiency comparison with SOTA. Finally, Section~VI concludes the paper.

%\vspace{-0.25em}
\section{Preliminaries: Arbiter PUFs}

A standard APUF launches two signal transitions into a cascade of challenge-controlled \texttt{MUX}s, as shown in Fig.~\ref{fig:figure2}(a). Each challenge bit determines whether the two signals propagate along the same path or cross to the opposite path. At the output, an arbiter (typically a latch) resolves which signal arrives first. Fabrication-induced process variations slightly perturb the delay of each path, and the sign of the cumulative delay difference produces a challenge-dependent response, forming a unique device fingerprint.

In conventional (\textit{deterministic}) PUF operation, challenges that result in small delay differences are typically avoided, as they lead to increased response instability. In contrast, a \textit{stochastic} APUF intentionally targets such near-balanced delay conditions. Under the influence of thermal noise and jitter, repeated evaluations of the same \textit{challenge} produce a probabilistic output, which can be modeled as a Bernoulli process with probability $p = P(S = 1)$.
%\begin{equation}
%p = P(S = 1).
%\end{equation}
This enables the same physical race mechanism to be repurposed as an SNG. %If two decorrelated stochastic bit-streams, \(S_A\) and \(S_B\), are generated, their associated probabilities can be manipulated to approximate arithmetic operations in the SC domain.
This perspective motivates the proposed architecture. Rather than treating small delay differences as a source of unreliability, they are leveraged as a design feature for low-overhead PUF-based stochastic bit-stream generation.

%\vspace{0.15em}
%\subsection{Formal Probabilistic Model}
\noindent\textbf{\textit{Formal Probabilistic Model.}}
Let the cumulative path delay difference at the arbiter input be $\Delta\! =\! T_{\text{Top}}\! -\! T_{\text{Bottom}}.$
%\begin{equation}
%\Delta = T_{\text{Top}} - T_{\text{Bottom}}.
%\end{equation}
If the arbiter decision is affected by a zero-mean random perturbation \(n\), then the effective decision variable becomes $z = f_{latch}(\Delta + n)$.
%\begin{equation}
%z = f_{latch}(\Delta + n)
%\end{equation}
The response bit (see Fig.~\ref{fig:figure2}(a)) is $S = 1$ if $z > 0$ and $S = 0$ otherwise.
%\begin{equation}
%S =
%\begin{cases}
%1, & z > 0 \\
%0, & z \le 0.
%\end{cases}
%\end{equation}
Thus, $P(S\!=\!1) \!=\! P(n\! >\! -\Delta).$
%\begin{equation}
%P(S=1) = P(n > -\Delta).
%\end{equation}
If \(n\) is Gaussian with variance \(\sigma_n^2\), then $P(S\!\!=\!\!1)\! =\! \Phi(\frac{\Delta}{\sigma_n})$,
%\begin{equation}
%P(S=1) = \Phi\!\left(\frac{\Delta}%{\sigma_n}\right),
%\end{equation}
where \(\Phi(\cdot)\) is the standard normal CDF. This simple model explains the key operating modes:
\textcolor{Navy}{\ding{202}} If \(|\Delta|\!\gg\! \sigma_n\), the output is repeatable %and almost deterministic
(i.e., deterministic). 
\textcolor{Navy}{\ding{203}} If \(|\Delta|\! \approx\! 0\), the output is near the stochastic boundary, and \(P(S\!=\!1)\) tends toward 0.5.
\textcolor{Navy}{\ding{204}} If \(\Delta\) is moderately positive or negative, the challenge maps to a tunable stochastic source with probability between 0 \& 1.
This probabilistic model, grounded in the CDF of the Gaussian 
noise distribution, directly connects challenge-dependent delay 
mismatch to observable output probability: a challenge that 
produces a delay difference $\Delta$ will generate a bit-stream 
with $P(S\!=\!1) = \Phi(\frac{\Delta}{\sigma_n})$. In practice, the 
challenge-search problem %therefore
reduces to finding a challenge 
$C$ whose static delay difference $\Delta$ matches the target 
value $\Phi^{-1}(p_{\mathrm{target}}) \cdot \sigma_n$, making 
the optimization over $|\Delta|$ the central design task in the 
proposed architecture.

%\vspace{-0.25em}
\section{Proposed Architecture}
%\vspace{-0.15em}
Fig.~\ref{fig:APUF_SNG} provides a high-level overview of the proposed \Design{} design flow, from post-fabrication characterization and challenge selection through to stochastic bit-stream generation for SC operations. Fig.~\ref{fig:figure2} complements this by illustrating the underlying physical mechanism. Specifically, Fig.~\ref{fig:figure2}(a) shows an $N$-stage APUF operating as a true-random bit-stream generator, where the arbiter decision is governed by the cumulative path delay difference $\Delta$. 
Fig.~\ref{fig:figure2}(b) shows how that decision becomes probabilistic: a fresh noise sample n is drawn at every evaluation and is centered on $\Delta$, so the fraction of evaluations resolving to `1' is the area of the noise density above the arbiter threshold, $P(S=1)=\Phi(\frac{\Delta}{\sigma_n})$. The mapping alongside it shows that selecting a challenge places $\frac{\Delta}{\sigma_n}$ at a chosen point on this curve; the stored vectors $C_{P1}$ and $C_{P2}$ of Fig.~\ref{fig:APUF_SNG} therefore yield the targets $p = 0.34$ and $p = 0.76$, rather than only the unbiased $p\approx\, 0.5$ obtained when $\Delta \approx\,0$. Fig.~\ref{fig:figure2}(c) shows the resulting SC bit-stream, whose probability of a `1' is set by the applied challenge vector.

A practical circuit implementation can be described as follows. Each stage of the APUF consists of a pair of crossable delay paths controlled by a challenge bit. These paths form a cascade of multiplexing elements that conditionally swap or preserve signal propagation paths based on the applied challenge vector. As the signal propagates through the stages, process variations and intentional or parasitic mismatches introduce cumulative delay differences between the two paths.
At the output, a timing comparator (arbiter) determines which signal arrives first, producing a binary response. The overall structure can be represented by: (i) a launch signal or clocked transition source, (ii) a cascade of challenge-controlled crossover stages, (iii) stage-dependent delay mismatches, (iv) a timing-based arbiter for signal comparison, and (v) an optional output register for synchronous sampling.

When the delay difference between the two paths is large, the arbiter produces a deterministic output. However, when the delay difference approaches zero, the decision becomes %highly 
sensitive to intrinsic noise sources such as thermal noise and local device variations. In this near-metastable regime, the output exhibits probabilistic behavior, enabling direct generation of stochastic bit-streams. This property forms the basis of the proposed architecture.

%\vspace{-0.25em}
\section{Analytical Modeling and Simulation}

\subsection{Arbiter PUF Modeling}
For an $N$-stage APUF, a signal race occurs between two symmetric \texttt{MUX} chains~\cite{Gao2020,Suh_DAC2007,Gassend_2002,Majzoobi_2008}.
Let the challenge vector applied to the PUF be denoted as $C\! \in\! \{0, 1\}^N$. The manufacturing variations are captured by a delay vector $\vec{w} \in \mathbb{R}^{N+1}$: one mismatch term per stage plus a final term for the asymmetry between the last stage and the arbiter~\cite{Gassend_2002, Herder_Tutorial_2014}. Each element is modeled as $w_i \!\sim\! \mathcal{N}(0, \sigma_{static}^2)$.
To determine the cumulative delay difference at the output arbiter, we define a parity vector $\vec{\Psi} \in \{-1, 1\}^{N+1}$, where each element represents the crossover state of the signal paths. The relationship between the binary challenge bits and the parity vector is given by $\Psi_i = \prod_{j=i}^{N} (1 - 2C_j)$ for $i = 1, \dots, N$, and $\Psi_{N+1} = 1$.
The total static delay difference between the top and bottom paths, $\Delta D_{static}$, is the linear combination of the delay vector and the parity vector. For a given chip and challenge, $\Delta D_{static}(C) = \vec{w} \cdot \vec{\Psi}(C) = \sum_{i=1}^{N+1} w_i \Psi_i$.
Because the delay components $w_i$ are independent and normally distributed, the total static delay across the state space of the PUF is distributed as $\Delta D_{static} \sim \mathcal{N}(0, (N+1)\sigma_{static}^2)$ (Here, we consider $\vec{\Psi}$ as the parity vector while $\Phi$ is reserved for the standard normal CDF).

\begin{algorithm}[!b]
%\vspace{-1em}
\caption{APUF-based Stochastic Bit-stream Generation}
\label{alg:puf_sc_gen_updated}
\footnotesize
%\linespread{0.8}\selectfont
\begin{algorithmic}[1]
\State \textbf{Inputs:} Target probability $P_{target}$, number of stages $N$, bit-stream length $L$, %bit-precision $p$, %max search iterations $M$, 
noise parameters $\sigma_{static}, \sigma_{dynamic}$
\State \textbf{Output:} Optimal challenge $\vec{C}_{opt}$, Stochastic bit-stream $\mathcal{B}$ of length $L$%$L = 2^p$

\Statex \textcolor{Navy}{\textbf{\hl{Step 1: Hardware Profiling}}}
\State $\vec{w} \leftarrow \text{randn}(N+1, 1) \cdot \sigma_{static}$ 
\Comment{\textcolor{DodgerBlue4}{Initialize static chip variations. simulation only; fitted from challenge-response pairs on silicon}}
%\hfill \textit{// Initialize static chip variations}
\State $\Delta D_{target} \leftarrow \text{norminv}(P_{target}, 0, \sigma_{dynamic})$ \Comment{\textcolor{DodgerBlue4}{Target delay for biased case}}

\Statex \textcolor{Navy}{\textbf{\hl{Step 2: Challenge Vector Search}}}
%\State $min\_err \leftarrow \infty, \quad best\_delay \leftarrow 0$
%\For{$i = 1$ to $M$}
%    \State $\vec{C} \leftarrow \text{randi}([0, 1], N, 1)$ \Comment{Generate random challenge}
    
%    \State \textit{// Calculate Parity Vector $\vec{\Phi}$ recursively}
%    \State $\Phi[N+1] \leftarrow 1$
%    \For{$j = N$ down to $1$}
 %       \State $\Phi[j] \leftarrow (1 - 2 \cdot C[j]) \cdot \Phi[j+1]$
%    \EndFor
    
%    \State $\Delta D_{static} \leftarrow \vec{w} \cdot \vec{\Phi}$ \Comment{Deterministic static delay difference}
%    \State $current\_err \leftarrow |\Delta D_{static} - \Delta D_{target}|$
    
%    \If{$current\_err < min\_err$}
%        \State $min\_err \leftarrow current\_err, \quad best\_delay \leftarrow \Delta D_{static}$
%    \EndIf
    
%    \State \textit{// Early exit if target bias is reached}
 %   \If{$min\_err < (\sigma_{dynamic} / 50)$}
%        \State \textbf{break}
%    \EndIf
%\EndFor

\State $min\_err \gets \infty$, %$best\_delay \gets 0$, 
$\vec{C}_{opt} \gets \varnothing$
\ForAll{$\vec{C} \in \{0,1\}^N$}
  \State $\Psi[N+1] \gets 1$ \Comment{\textcolor{DodgerBlue4}{$\Psi$: Parity vector}}
  \For{$j = N$ \textbf{down to} $1$}
    \State $\Psi[j] \gets (1 - 2\cdot C[j]) \cdot \Psi[j+1]$
  \EndFor
  \State $\Delta D_{static} \gets \vec{w} \cdot \vec{\Psi}$ \Comment{\textcolor{DodgerBlue4}{Static delay difference}}
  \State $current\_err \gets |\Delta D_{static} - \Delta D_{target}|$
  \If{$current\_err < min\_err$}
    \State $min\_err\! \gets\! current\_err$, $best\_delay\! \gets\! \Delta D_{static}$, $\vec{C}_{opt}\! \gets\! \vec{C}$
  \EndIf
\EndFor

\Statex \textcolor{Navy}{\textbf{\hl{Step 3: Bit-stream Evaluation}}}
\For{$k = 1$ to $L$}
    \State $N_{th} \leftarrow \text{randn}() \cdot \sigma_{dynamic}$ \Comment{\textcolor{DodgerBlue4}{Sample dynamic thermal noise}}
    \State $\Delta D_{total} \leftarrow best\_delay + N_{th}$
    
    \If{$\Delta D_{total} > 0$}
        \State $\mathcal{B}[k] \leftarrow 1$
    \Else
        \State $\mathcal{B}[k] \leftarrow 0$
    \EndIf
\EndFor
\State \Return $\mathcal{B}$, $\vec{C}_{opt}$
\end{algorithmic}
\end{algorithm}

\subsection{Controlling Probability %Stochastic Probability Generation 
via Thermal Noise}
To transition the PUF from  deterministic operation to true-random bit-stream generation, %TRNG suitable for SC, 
we must exploit dynamic evaluation noise. At runtime, thermal noise and power supply jitter introduce instantaneous variations to the signal propagation time. We model this dynamic noise as an additive random variable $N_{dynamic}$ drawn from normal distribution %(with mean 0 and variance of $\sigma_{dynamic}^2$) 
as $N_{dynamic} \sim \mathcal{N}(0, \sigma_{dynamic}^2)$.
Therefore, the instantaneous delay difference at the arbiter for a single evaluation is: $ \Delta D_{total} = \Delta D_{static}(C) + N_{dynamic}.$
%\begin{equation}
%    \Delta D_{total} = \Delta D_{static}(C) + N_{dynamic}.
%\end{equation}
The arbiter resolves the race condition to produce a binary response output $S\! \in\! \{0, 1\}$. Assuming the arbiter triggers a `1' when the top path arrives first ($\Delta D_{total} > 0$), the probability of generating a `1', denoted as $p$, is the probability that the thermal noise pushes the total delay above zero: $p\! =\! P(S\! =\! 1)\! =\! P(\Delta D_{static}(C)\! +\! N_{dynamic}\! >\! 0).$
%\begin{equation}
%    p = P(Y = 1) = P(\Delta D_{static}(C) + N_{dynamic} > 0).
%\end{equation}
Given the symmetrical nature of the normal distribution, we can map \textit{any target probability} $p_{target}$ to a specific required static delay with the inverse CDF of the standard normal distribution, $\Phi^{-1}$, as $\Delta D_{target} = \sigma_{dynamic} \cdot \Phi^{-1}(p_{target})$.
%\begin{equation}
%    \Delta D_{target} = \sigma_{dynamic} \cdot \Psi^{-1}(p_{target}).
%\end{equation}

%\color{black}

\subsection{Simulation Methodology}
The challenge space is finite. Evaluating one challenge costs $\mathcal{O}(N)$, so for $N \!\le\! 16$ we enumerate all $2^N$ challenges (65{,}536 for $N=16$, about $1.1\times10^6$ multiply-accumulates) and keep the one whose static delay is closest to $\Delta D_{target}$: $\vec{C}_{opt}\! =\! \operatorname*{argmin}_{C} |\Delta D_{static}(C)\! -\! \Delta D_{target}|$ (refer to Algorithm~\ref{alg:puf_sc_gen_updated}, and $\vec{C}_{opt}$ is the $N$-bit vector stored for that probability level). Enumeration is exact, so coverage rather than search decides accuracy. The $2^N$ delays follow $\mathcal{N}(0,(N+1)\sigma_{static}^2)$, and over $p \in [0.05, 0.95]$ the largest gap between achievable probabilities is 3\% for $N=8$ and 0.03\% for $N=16$. The latter is far below the 1\% sampling floor of a 1024-bit stream, which is why accuracy saturates at 16 stages.

%The simulation evaluates the viability of discovering challenges that satisfy $\Delta D_{target}$. Due to the NP-hard nature of reversing the PUF delay model analytically, we employ a Monte Carlo random search methodology, as shown in Algorithm~\ref{alg:puf_sc_gen_updated}. For a targeted SC bit-stream probability $p_{target}$, the algorithm proceeds as follows:

%\noindent \ding{172} \textbf{Hardware Profiling:} A static hardware instance is initialized by sampling $w_i$ for all $N$ stages, locking the $\mathcal{N}(0, N\sigma_{static}^2)$ distribution.

%\noindent \ding{173} \textbf{Challenge Search:} A finite set of $M$ random challenges is generated and evaluated. For each challenge $C_k$, the static delay $\Delta D_{static}(C_k)$ is calculated.

%\noindent \ding{174} \textbf{Error Minimization:} The algorithm selects the optimal challenge $C_{opt}$ that minimizes the absolute distance to the target delay: $C_{opt} = \underset{C_k}{\mathrm{argmin}} \left| \Delta D_{static}(C_k) - \Delta D_{target}\right|.$ 

%\noindent To optimize the calibration phase, an early-termination criterion was established at $\epsilon < \sigma_{dynamic} / 50$. This threshold ensures that the residual search error remains an order of magnitude below the dynamic noise floor, thereby preventing unnecessary computational overhead while maintaining a target probability accuracy within $\approx 1\%$.

%\noindent \ding{175} 
\textbf{Bit-stream Generation:} The selected challenge $\vec{C}_{opt}$ is evaluated $L$ times (where $L$ is the bit-stream length), with independent dynamic noise $N_{dynamic}$ injected in each cycle, yielding the final bit-stream.
To evaluate the quality of the generated bit-streams, we conduct Monte Carlo simulations with 10,000 iterations to convert the random numbers in the $[0,1]$ interval to their corresponding bit-streams and measure the mean absolute error (MAE). 
We evaluate the correlation level between pairs of generated bit-streams using the Stochastic Cross Correlation (SCC) metric~\cite{Alaghi_SCC1}:
%\vspace{-0.5em}
\begin{equation}
\label{SCC_equation}
    SCC = \begin{cases} \frac{ad-bc}{L \times min(a+b, a+c)-(a+b) \times(a+c)} & ,  \  if \ ad>bc \\ \frac{ad-bc}{(a+b) \times(a+c) - L \times max(a-d, 0)} & ,  \ else\end{cases}
\end{equation}
where, $a$, $b$, $c$, and $d$ represent the logic pairs $11$, $10$, $01$, and $00$, respectively, taken from the corresponding bit positions of the two bit-streams.
The level of correlation is determined by the number of overlapping logic-1s across all bit positions.
When this overlap is maximized, the bit-streams exhibit 
\textit{maximum correlation} ($SCC\!=\!+1$). When the overlap is minimized, they are in \textit{minimum correlation} ($SCC\!=\!-1$).
A particularly desirable case is \textit{zero correlation} ($SCC\!=\!0$), which is required for SC operations such as multiplication using bit-wise \texttt{AND} operation. 

\begin{figure*}[!t]
\vspace{-0.5em}
\centering
\begin{tabular}{ccc}
\includegraphics[width=0.25\linewidth]{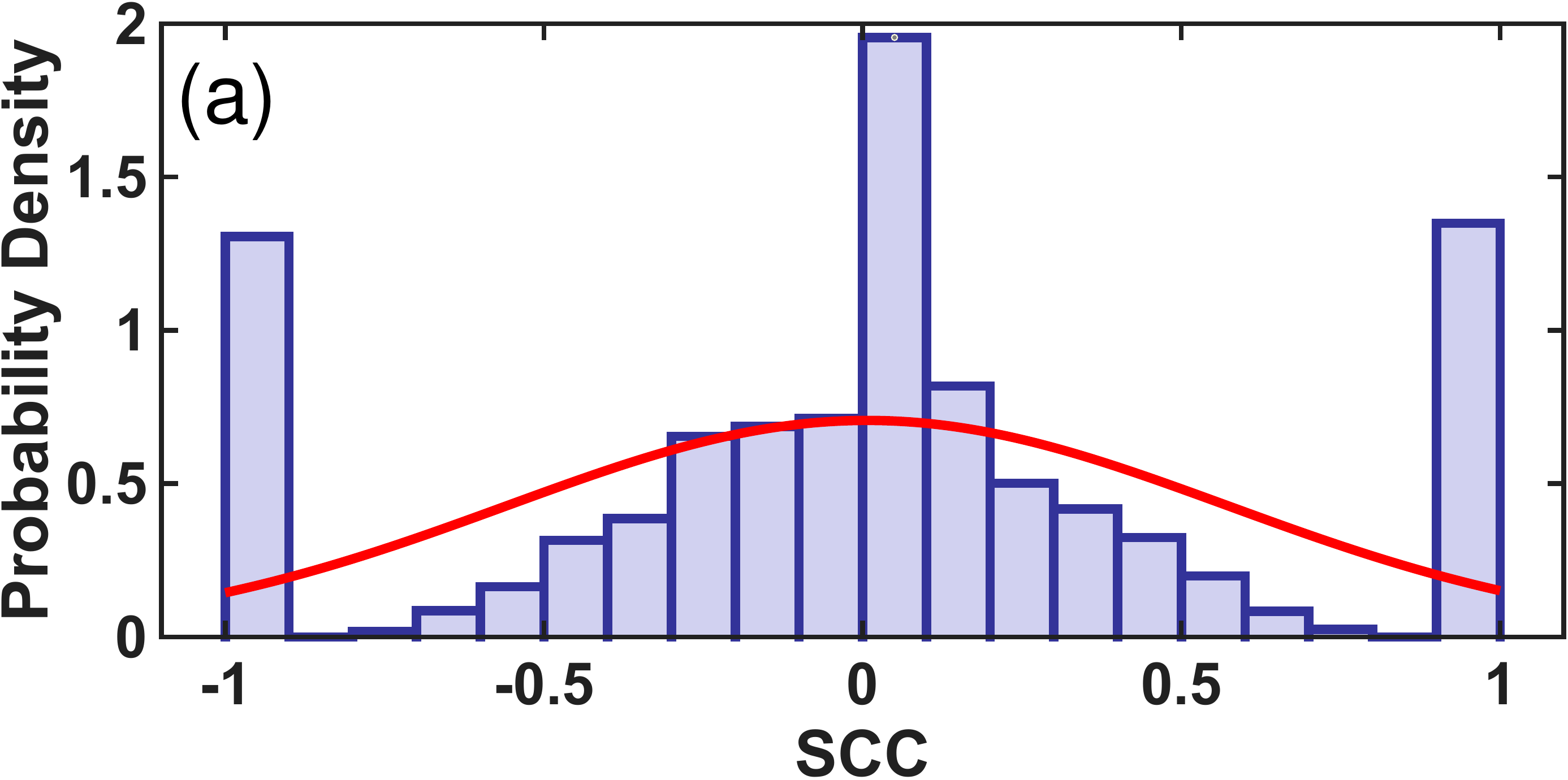} &
\includegraphics[width=0.25\linewidth]{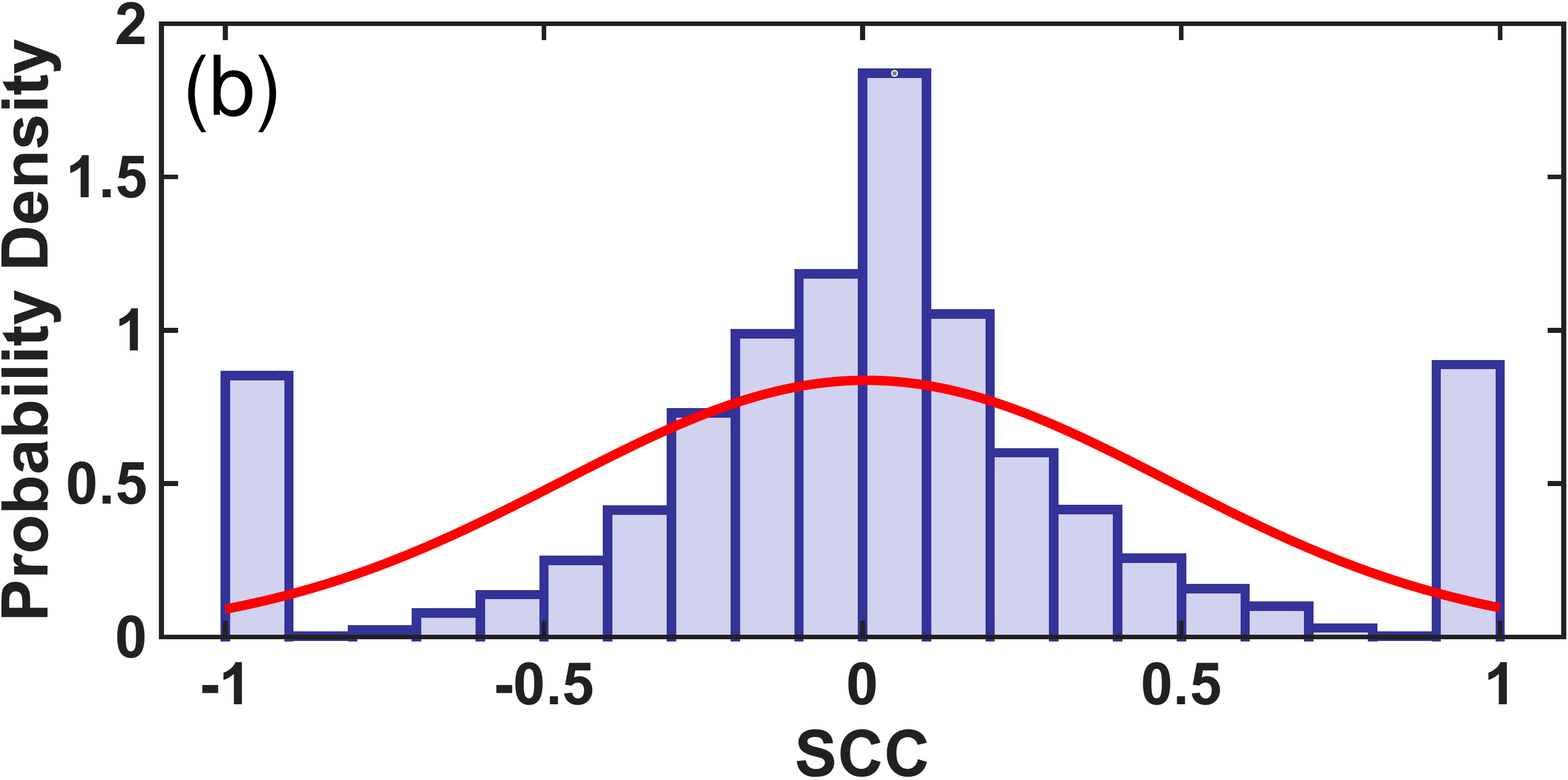} &
\includegraphics[width=0.25\linewidth]{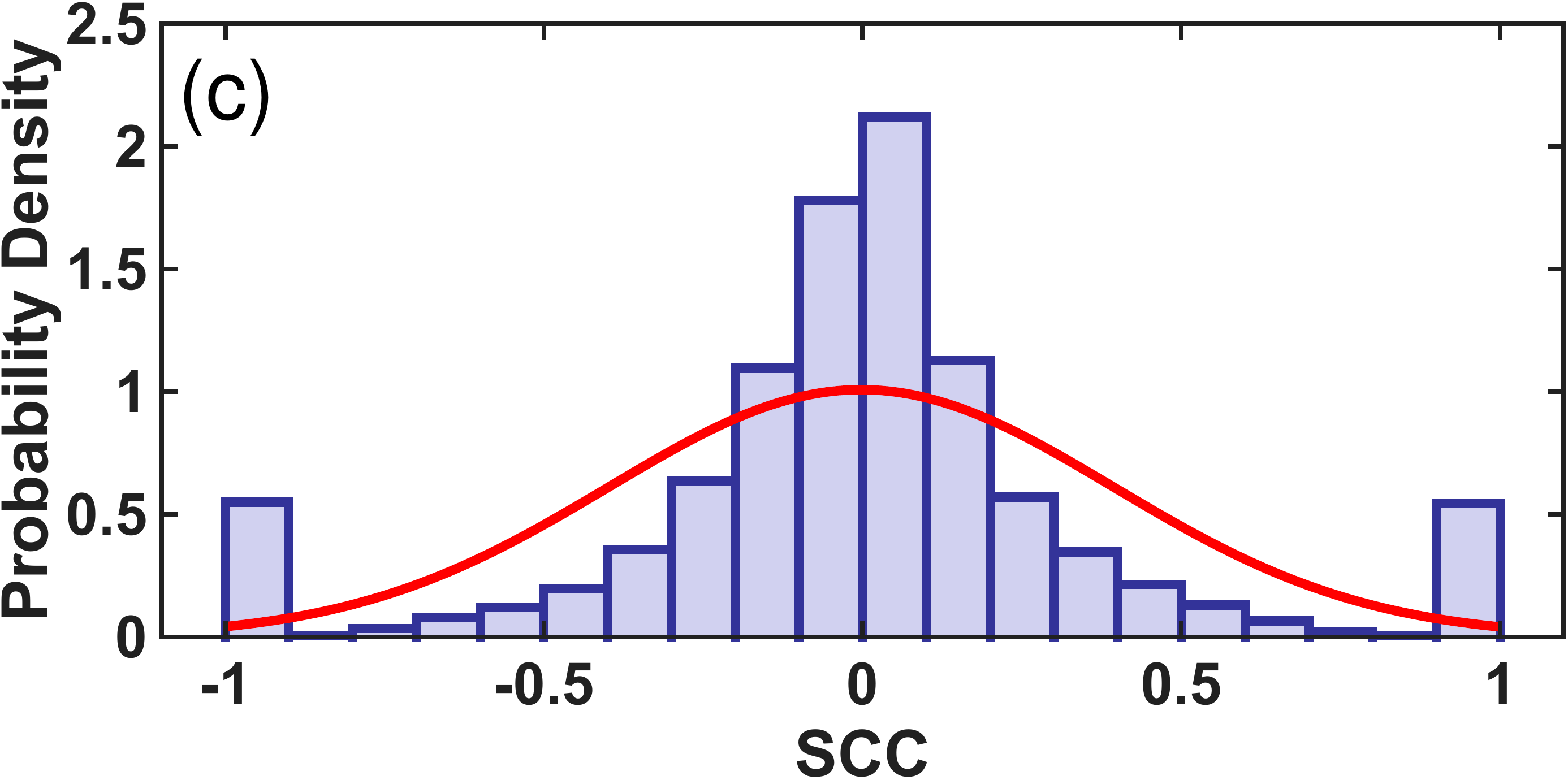} \\

\includegraphics[width=0.25\linewidth]{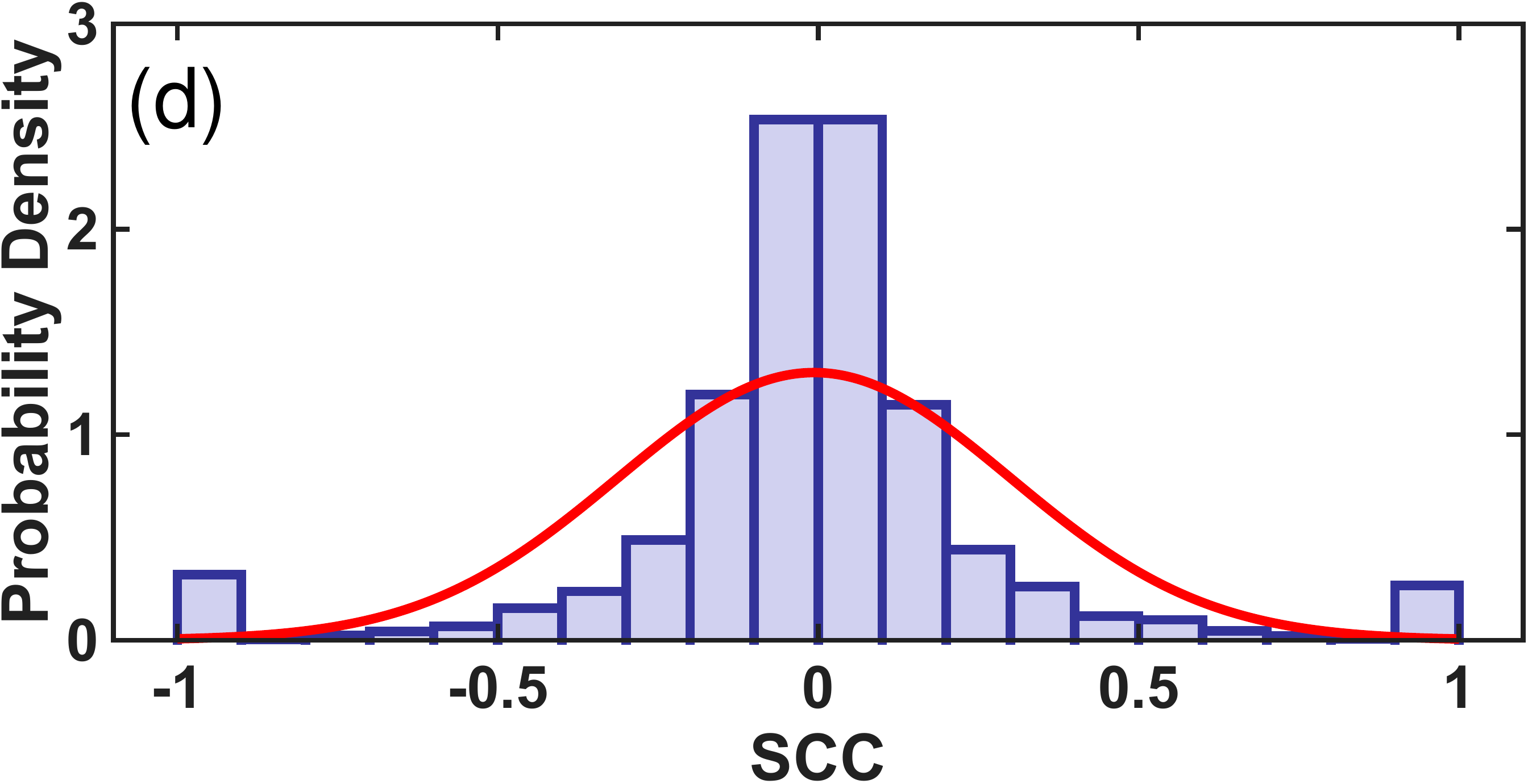} &
\includegraphics[width=0.25\linewidth]{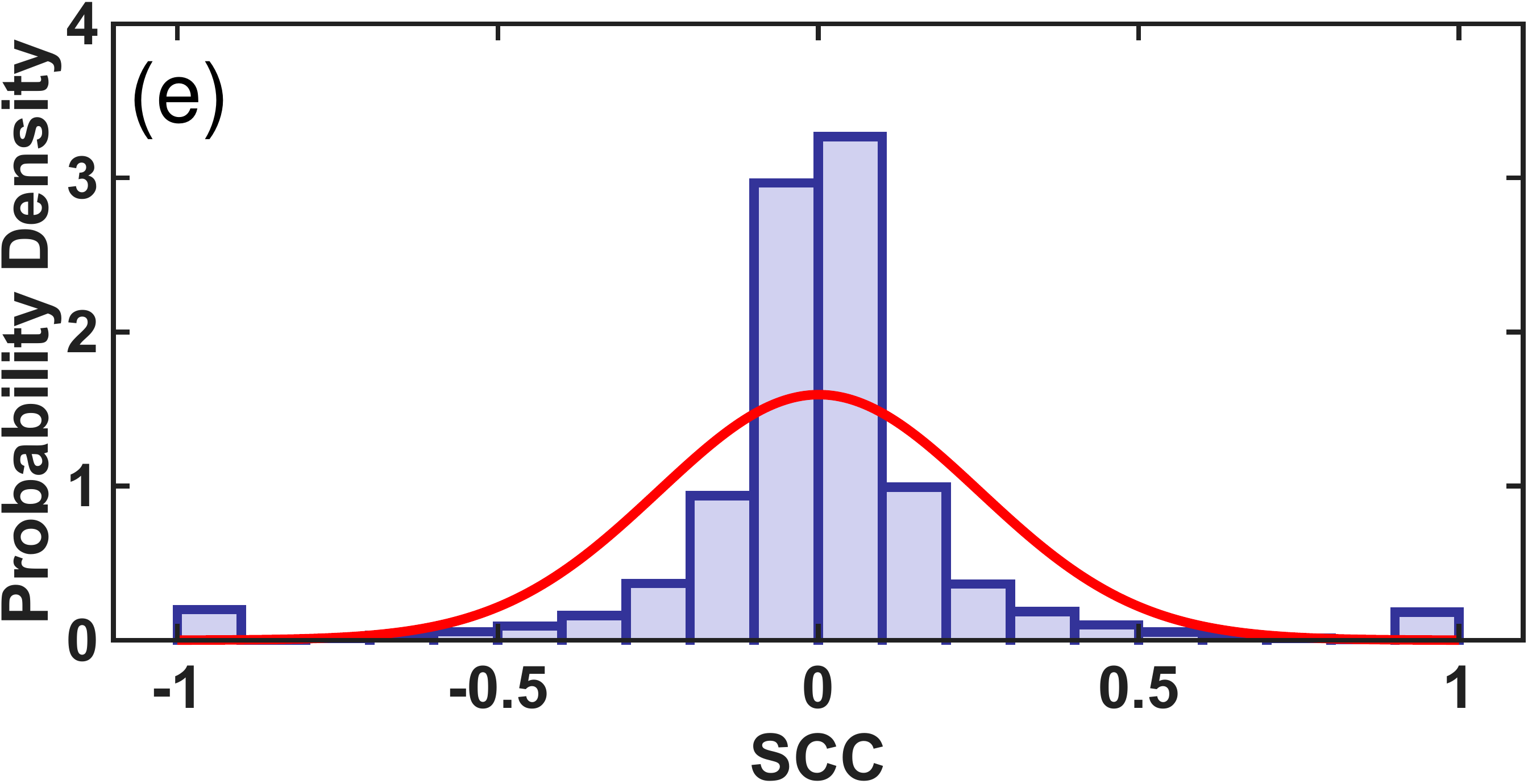} &
\includegraphics[width=0.25\linewidth]{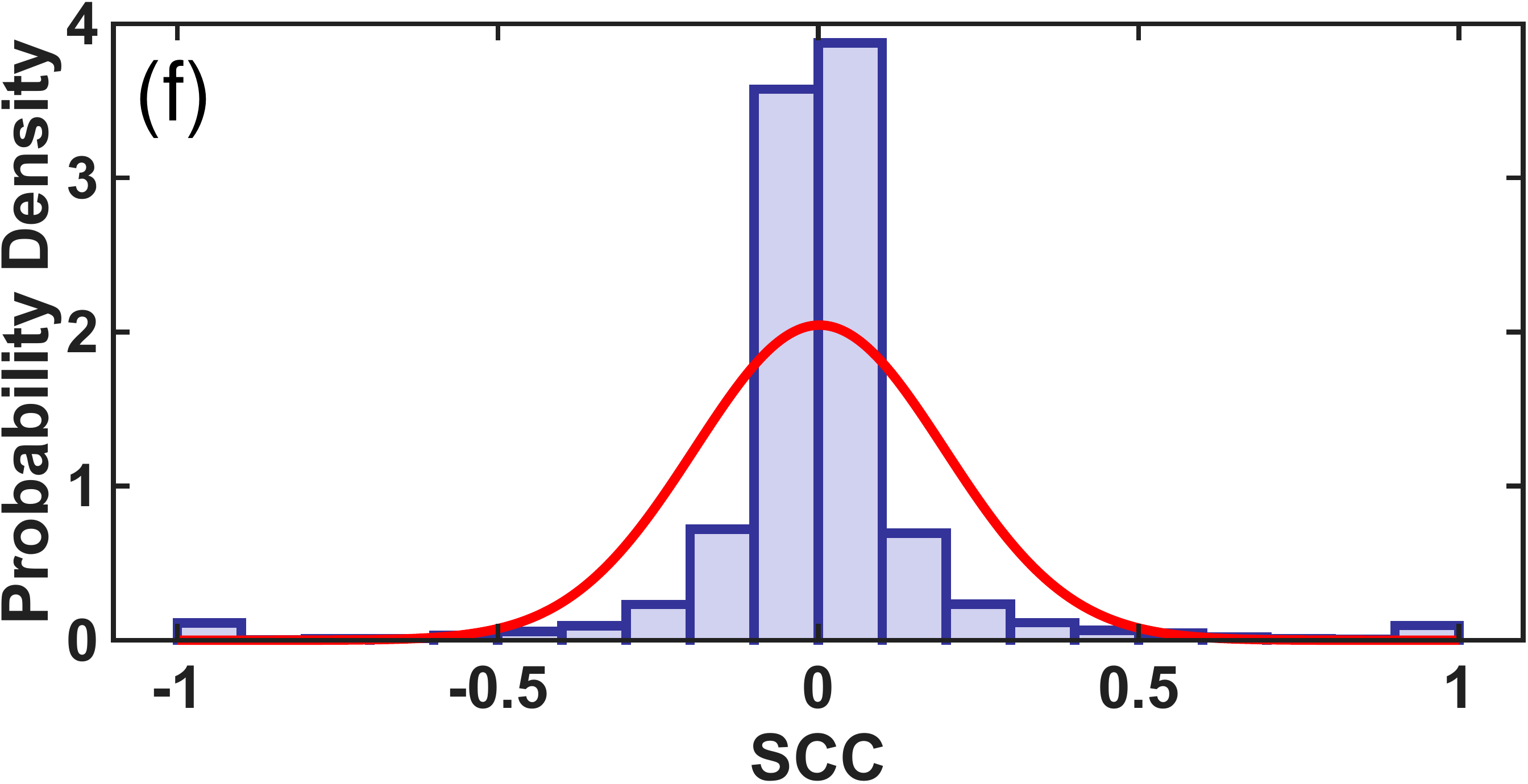}
\end{tabular}

\vspace{-0.5em}
\caption{SCC distribution between pairs of generated SC bit-streams using \Design~16-stage with bit-stream length of (a) 32-bits, (b) 64-bits, (c) 128-bits, (d) 256-bits, (e) 512-bits, and (f) 1024-bits. %Increasing the bit-stream length improves independency between bit-streams.
} 
\label{fig:scc_designs}
\vspace{-1.0em}
\end{figure*}

%To ensure the validity of the proposed \Design~for SC core computation logic, it is essential to verify the spatial independence of the generated bit-streams. To evaluate this, 

%We computed the SCC across thousands of independently generated bit-stream pairs. Fig.~\ref{fig:scc_designs}  illustrates the resulting histogram of the measured SCC values. To characterize the statistical profile of the spatial overlap, a normal distribution was fitted to the empirical histogram data. The fitted Gaussian curve demonstrates a mean ($\mu$) tightly centered at zero, confirming an absence of structural bias or spatial dependency between the generated bit-streams. As can be seen, increasing the bit-stream length provides more independent bit-streams as the variance becomes narrower accordingly.

%To verify the spatial independence of the generated bit-streams, we conducted an extensive SCC measurements with 10,000 iterations between each pair of 
%Figure~\ref{fig:scc_designs} illustrates the SCC distribution measurements between any generated pairs of bit-streams using an \Design~16-stage design for different bit-stream lengths of 32, 64, 128, 256, 512, and 1024.

Low correlation is structural in \Design. Each bit is $\mathbf{1}\{\Delta D_{static}(C)\!+\!n_k\!>\!0\}$ with a fresh noise sample $n_k\! \sim\! \mathcal{N}(0,\sigma_{dynamic}^2)$ per evaluation, so the challenge fixes only the mean and two streams have zero covariance whatever paths their challenges share. Fig.~\ref{fig:scc_designs} confirms this for random pairs, with a Gaussian fit centered at zero and spread $\frac{1}{\sqrt{L}}$. The harder cases behave the same: pairs mapped to the same target probability, identical challenges sharing every path (standard deviation 0.031 at $L$=$1024$, matching $\frac{1}{\sqrt{L}}$), and eight streams generated simultaneously from the same PUF instance, as an SC datapath would, %and eight streams generated together as in an SC datapath, 
whose 28 pairwise SCCs remain centered at zero.

%==================================================================
%    Mehran - Accuracy Table for SC bit-stream generation quality
%==================================================================
\begin{table}[!b]
\centering
\vspace{-1em}
\caption{MAE (\%) Comparison of SC Bit-stream Generation Methods}
\vspace{-.5em}
\renewcommand{\arraystretch}{0.3}
% Use the D column type for decimal alignment
%\begin{tabular}{l l *{8}{D{.}{.}{2.3}}}
\resizebox{\columnwidth}{!}{
\begin{tabular}{l *{7}{c}}
\toprule
\multirow{2}{*}{\textbf{Method}} & \multicolumn{7}{c}{\textbf{Bit-stream length (L)}} \\
\cmidrule(lr){2-8}
% Wrap each header number in \multicolumn to center it
& \multicolumn{1}{c}{\textbf{16}} & \multicolumn{1}{c}{\textbf{32}} & \multicolumn{1}{c}{\textbf{64}} & \multicolumn{1}{c}{\textbf{128}} & \multicolumn{1}{c}{\textbf{256}} & \multicolumn{1}{c}{\textbf{512}} & \multicolumn{1}{c}{\textbf{1024}} \\
\midrule

\Design--4 stages & 10.74 & 6.39 & 4.59 & 4.24 & 3.17 & 9.45 & 5.24 \\
						
\Design--5 stages & \cellcolor{gray!15}8.25 & \cellcolor{gray!15}6.15 & \cellcolor{gray!15}4.57 & \cellcolor{gray!15}3.96 & \cellcolor{gray!15}3.00 & \cellcolor{gray!15}3.70 & \cellcolor{gray!15}2.27 \\
						
\Design--6 stages & 8.41 & 6.27 & 4.41 & 2.90 & 2.48 & 2.60 & 1.56 \\
						
\Design--7 stages & \cellcolor{gray!15}7.97 & \cellcolor{gray!15}5.78 & \cellcolor{gray!15}3.91 & \cellcolor{gray!15}2.77 & \cellcolor{gray!15}2.03 & \cellcolor{gray!15}2.67 & \cellcolor{gray!15}1.93 \\
						
\Design--8 stages & 7.84 & 6.16 & 3.91 & 2.80 & 1.96 & 1.59 & 1.04 \\
						
\Design--16 stages & \cellcolor{gray!15}7.90 & \cellcolor{gray!15}5.48 & \cellcolor{gray!15}3.95 & \cellcolor{gray!15}2.76 & \cellcolor{gray!15}1.95 & \cellcolor{gray!15}1.42 & \cellcolor{gray!15}1.01 \\
						
\Design--32 stages & 7.84 & 5.57 & 3.90 & 2.75 & 1.98 & 1.41 & 1.02 \\
						
\Design--64 stages & \cellcolor{gray!15}7.86 & \cellcolor{gray!15}5.63 & \cellcolor{gray!15}3.94 & \cellcolor{gray!15}2.76 & \cellcolor{gray!15}1.96 & \cellcolor{gray!15}1.39 & \cellcolor{gray!15}1.01 \\
						
\Design--128 stages & 7.82 & 5.50 & 3.95 & 2.74 & 1.97 & 1.42 & 1.01 \\

\midrule
% Final data rows
\texttt{MATLAB} (rand) &\cellcolor{gray!15}7.84 &\cellcolor{gray!15}5.54 &\cellcolor{gray!15}3.91 &\cellcolor{gray!15}2.76 &\cellcolor{gray!15}1.96 &\cellcolor{gray!15}1.38 &\cellcolor{gray!15}0.97 \\
PRNG ($\log_2{L}$-bit LFSR)\textsuperscript{$\ast$} & 6.10 & 3.17 & 1.51 & 0.77 & 0.39 & 0.21 & 0.10 \\
QRNG (Sobol) & \cellcolor{gray!15}3.13 &\cellcolor{gray!15}1.56 &\cellcolor{gray!15}0.78 &\cellcolor{gray!15}0.39 &\cellcolor{gray!15}0.19 &\cellcolor{gray!15}0.09 &\cellcolor{gray!15}0.04 \\
TRUE-BSG~\cite{TRUE_BSG_ISCAS2025}\textsuperscript{\dag} & 7.94 & 5.62 & 3.97 & 2.77 & 1.97 & 1.37 & 1.06 \\
%TRNG--ReRAM
IMSNG~\cite{SC_IMC_DAC2025}\textsuperscript{\ddag} & \cellcolor{gray!15}7.90 & \cellcolor{gray!15}5.55 & \cellcolor{gray!15}4.08 & \cellcolor{gray!15}3.12 & \cellcolor{gray!15}2.44 & \cellcolor{gray!15}2.01 & \cellcolor{gray!15}1.78 \\
\bottomrule
%\multicolumn{10}{l}{\scriptsize For PRNG, we use a maximal length LFSR with polynomial $x^8+x^5+x^3+1$.} % Table note, left-aligned
\end{tabular}
}
\parbox{\columnwidth}{\scriptsize
$\ast$ maximal-period LFSR. \ \dag, \ddag~Optimal cases were selected. 
}
\label{tab:accuracy_bs}
%\vspace{-1em}
\end{table}

%\vspace{-0.5em}
\section{Evaluation Results and Discussion}
%\vspace{-0.3em}

%This section evaluates the performance of the proposed \Design~in generating SC bit-streams with desired probabilities and in performing SC arithmetic operations on the generated bit-streams.
The evaluation builds on the additive delay model, which is the standard APUF model and has been validated on fabricated devices~\cite{Gassend_2002, Herder_Tutorial_2014, Suh_DAC2007}, and on the near-metastable randomness mechanism, which prior PUF-TRNG designs have measured in silicon~\cite{11421416, 10150018, 9935279}. On hardware, the challenge table is built from each challenge's measured output probability rather than from the model, so the accuracy reported below does not depend on the model being exact (Section~\ref{sec:accuracy}). What silicon adds is the size of voltage, temperature, and aging drift, which Section~\ref{sec:stability} bounds.
Throughout all experiments, $\sigma_{static} = 0.03$ and $\sigma_{dynamic} = 0.07$. Designs with $N \le 16$ use exhaustive enumeration, and the 32 to 128-stage designs use 10{,}000 random challenges.
%Throughout all experiments, Algorithm~\ref{alg:puf_sc_gen_updated} is configured with  %We initialized the input parameters of 
%\textit{max search iterations ($M$)} = 10,000, \textit{static noise standard deviation ($\sigma_{static}$)} = 0.03, and \textit{dynamic noise standard deviation ($\sigma_{dynamic}$)} = 0.07. 
%in Algorithm~\ref{alg:puf_sc_gen_updated}. 
{%Increasing the $\sigma_{dynamic}$ may collapse the quality of the generated bit-streams and accordingly affects the arithmetic operations accuracy. So it needs to be smartly selected. Usually we use small values. 
%0.7 is not a good choice.
The dynamic noise parameter $\sigma_{\mathrm{dynamic}}$ must be 
chosen carefully: too large a value flattens the 
$\Phi(\frac{\Delta}{\sigma_{\mathrm{dynamic}}})$ curve, reducing 
probability resolution and degrading both bit-stream quality 
and arithmetic accuracy, while too small a value makes the 
challenge search overly sensitive to quantization errors in 
$\Delta$. The selected value of $\sigma_{\mathrm{dynamic}} = 0.07$ 
balances these competing effects.

%\vspace{-0.5em}
\subsection{Accuracy Evaluation}
\label{sec:accuracy}
%\vspace{-0.2em}
\subsubsection{\textbf{Bit-stream generation}} We first evaluate the proposed \Design~design in generating SC bit-streams compared to SOTA PRNG- and QRNG-based bit-stream generation methods, as well as a software generator that utilizes the \texttt{MATLAB} \texttt{rand} function as the RNG. For each method, we convert 10,000 random real numbers in the $[0,1]$ interval into SC bit-streams. Each input value is converted to SC bit-streams of various lengths ({$\forall\, L\!\in\! \{16,32,64,128,256,512,1024\}$}). For optimal performance of the PRNG method, we implemented maximal-period LFSRs sized according to the bit-stream length ($L$). %, i.e., a $\log_2{L}$-bit maximal-period LFSR is used to generate an $L$-bit bit-stream. 
For the QRNG method, we used the first Sobol sequence from \texttt{MATLAB}'s Sobol source generator~\cite{Siting_Sobol_TVLSI2018}.
Table~\ref{tab:accuracy_bs} reports the MAE between the target input values and the values represented by the generated bit-streams.
%For each case, we measure the MAE by calculating the absolute difference between the target input value and the value represented by the generated  bit-stream (Table~\ref{tab:accuracy_bs}).
For \Design~with %configurations utilizing 
4 to 7 stages, the small state space limits accuracy.
%the mathematical limitations of a constrained state space are visible.
The \Design--4 stage design exhibits severe quantization penalties, failing to accurately synthesize fine-grained target probabilities. This results in inconsistent error scaling and a high MAE of 5.24\% at a bit-stream length of 1024. The coarse delay resolution of these shallow designs prevents the hardware from effectively mitigating dynamic thermal noise fluctuations.
As the number of stages increases to 8, the MAE stabilizes. %drastically.
%Essentially, 
The \Design--16 stage design achieves an MAE of 1.01\% at $L=1024$. 
The 32 to 128-stage designs match within 1.01\%, so 16 stages saturate accuracy.
%This performance is statistically identical to the 32, 64, and 128-stage designs. This empirical convergence proves that 16 stages saturate the accuracy ceiling. 
%As an optimum point, 
%\Design--8 design provides the optimal architectural balance with high accuracy while eliminating the routing complexity, dynamic power consumption, and silicon area overhead associated with deeper \texttt{MUX} chains. These measurements validate the \Design--8 stage as a reliable SNG providing comparable accuracy to the software SNG. 

Compared to the SOTA TRNG-based SC bit-stream generators, the \Design--8 stage matches the accuracy of TRUE-BSG design %standard Ring Oscillator TRNGs 
(1.06\%)~\cite{TRUE_BSG_ISCAS2025} while outperforming the true-random generator developed with emerging non-volatile ReRAM devices, IMSNG (1.78\%)~\cite{SC_IMC_DAC2025}. % demonstrating its effectiveness in generating %converting unbiased random bits into 
%high-quality SC bit-streams.
%Finally, considering the synthetic randomness, 
%The evaluation demonstrates that 
The PRNG (LFSR) and QRNG (Sobol sequence) approaches achieve the lowest MAE, reaching 0.10\% and 0.04\% at 1024 bits, respectively. 
This is expected: %This discrepancy is mathematically expected. 
utilizing Quasi-Randomness does not require excessive runs due to its inherent deterministic nature. Sobol sequences guarantee uniform, space-filling probability distributions that prevent the random bit-clustering (fluctuation error) unavoidable in pure thermal-noise-driven TRNGs. 
This floor does not depend on the noise law. With the table calibrated to the actual noise, MAE at $L\!=\!1024$ stays between 0.97\% and 0.99\% for Gaussian, Laplace, uniform, and Student-$t$ noise, and for $\sigma_{dynamic}$ anywhere from 0.04 to 0.15. Only open-loop calibration is sensitive: assuming Gaussian noise when the true noise is heavy-tailed raises MAE to 5.6\%. On silicon the table comes from each challenge's measured probability, so this error does not arise.
%However, 
While QRNGs provide peak computational accuracy for SC, they demand significant hardware overhead. The proposed 8-stage \Design~sacrifices this fractional deterministic accuracy to provide a lightweight hardware footprint paired with physical unclonability and inherent device security.

%\noindent\Mehcolor{
\subsubsection{\textbf{Arithmetic operations}}
Table~\ref{tab:sc_arithmetic_comparison} expands the accuracy evaluation from raw bit-stream generation to the computational accuracy of basic arithmetic operations. The \textit{SC Arithmetic Operations} panel of Fig.~\ref{fig:APUF_SNG} shows how essential arithmetic operations are implemented in SC with simple logic elements.
For arithmetic operations that demand correlated bit-streams at their inputs (e.g., \textit{minimum}, \textit{maximum}, \textit{division}, and \textit{absolute value subtraction}), we employed the correlation manipulation technique proposed in~\cite{asadi2025eco} to increase the correlation level.
This adds only combinational logic at the operation inputs and does not change the generator itself.
%\newline
The results illustrate how the probability quantization errors inherent to shallow APUF configurations (4 to 7 stages) cascade through combinational and sequential SC logic. In the 4-stage design, operations sensitive to bit-stream correlation and precision (e.g., division and absolute subtraction) exhibit severe degradation. For instance, the division operation yields an inconsistent MAE of 24.48\% at 128 bits, while absolute subtraction peaks at 17.13\%. %The coarse delay resolution of these configurations leads to fluctuation and inconsistency in accuracy measurements.
Consistent with the raw generation metrics, the MAE stabilizes at 8 stages and converges at 16 stages. For bit-stream length of 1024, the \Design--16 stage design achieves low errors: 0.89\% (multiplication), 1.03\% (scaled addition), 1.47\% (abs. subtraction), 1.01\% (minimum), and 1.03\% (maximum). Scaling the APUF state space to 32, 64, or 128 stages provides no benefit across any operation (e.g., multiplication remains locked at 0.89\% to 0.90\%). %The accuracy results suggests $N=8$ architecture as the optimal threshold, providing accurate arithmetic without the high area and power overheads of deeper \texttt{MUX} chains.

The \Design--8 stage design tracks %proves highly competitive against the software generator, %ideal software bounds, 
%closely mirroring 
the \texttt{MATLAB} (rand) baseline across all arithmetic operations.
Notably, this design %the \Design--16 stage significantly 
outperforms the LFSR in sequential logic operations like \textit{division} operation. While the LFSR-based design achieves a lower MAE in basic \textit{multiplication} (0.48\% at 1024 bits), its inherent temporal autocorrelation completely breaks down in the \textit{division} circuit, resulting in a high 8.39\% error. %The \Design--8 stage %, driven by true thermal noise, 
%maintains \red{statistical independence} over time, reducing the division error to 3.54\%.
While the QRNG design dominates all randomness sources (pushing \textit{multiplication} MAE down to 0.07\% at 1024 bits) this is expected due to the deterministic, uniform spatial distribution of LD sequences. However, deploying QRNGs requires complex, high-overhead hardware sequence generators and lacks the benefits of true randomness.

%When benchmarked against alternative hardware true randomness sources, the \Design--8 design demonstrates strict equivalency or superiority. It provides comparable accuracy to TRUE-BSG~\cite{TRUE_BSG_ISCAS2025} across various arithmetic operations (e.g., 0.88\% vs 0.91\% for multiplication) while outperforms the ReRAM TRNG approach~\cite{SC_IMC_DAC2025} (1.45\% for multiplication and 2.20\% for abs. subtraction). The 8-stage \Design~delivers the computational robustness of an ideal thermal TRNG, avoids the autocorrelation pitfalls of LFSRs in sequential SC elements, and achieves this with a silicon footprint significantly smaller than competing QRNG-based generators. 

%==================================================================
%       Mehran - Accuracy Table for SC Arithmetic operations
%==================================================================
\begin{table}[!t]
\centering
\caption{MAE (\%) Comparison of SC Arithmetic Operations using Different Randomness Sources}
% Set up the S column type for numbers with max 2 digits before and 2 after the decimal.
%\sisetup{table-format=2.2}
\vspace{-0.75em}
\renewcommand{\arraystretch}{0.7}
\resizebox{\columnwidth}{!}{
\begin{tabular}{ll *{7}{c}}
\toprule
% Table Header
\multirow{2}{*}{\begin{tabular}[c]{@{}l@{}}\textbf{Method} \end{tabular}} & \multirow{2}{*}{\textbf{\begin{tabular}[c]{@{}l@{}}SC Arithmetic \\ Operation\end{tabular}}} & \multicolumn{7}{c}{\textbf{Bit-stream length (L)}} \\
\cmidrule(lr){3-9} % Partial rule under the main header
& & \textbf{16} & \textbf{32} & \textbf{64} & \textbf{128} & \textbf{256} & \textbf{512} & \textbf{1024} \\
\midrule

% Data Section 6
\multirow{7}{*}{\begin{tabular}[c]{@{}l@{}} \Design \\(4 stages) \end{tabular}}
&\cellcolor{gray!15}Multiplication &\cellcolor{gray!15}8.74 &\cellcolor{gray!15}5.51 &\cellcolor{gray!15}3.86 &\cellcolor{gray!15}3.59 &\cellcolor{gray!15}2.58 &\cellcolor{gray!15}7.15 &\cellcolor{gray!15}4.15 \\
& Scaled Addition & 8.80 & 6.54 & 5.06 & 8.83 &	6.10 & 10.94 & 3.40 \\
&\cellcolor{gray!15}Abs. Subtraction &\cellcolor{gray!15}13.90 &\cellcolor{gray!15}11.59 &\cellcolor{gray!15}9.15 &\cellcolor{gray!15}17.13 &\cellcolor{gray!15}11.51 &\cellcolor{gray!15}16.65 &\cellcolor{gray!15}3.98 \\
& Division & 19.67 & 18.2 &	14.87 &	24.48 & 19.02 &	17.59 &	6.90 \\
&\cellcolor{gray!15}Minimum&\cellcolor{gray!15}8.78 &\cellcolor{gray!15}8.61 &\cellcolor{gray!15}6.80 &\cellcolor{gray!15}11.37 &\cellcolor{gray!15}9.73 &\cellcolor{gray!15}7.05 &\cellcolor{gray!15}3.03 \\
& Maximum & 11.72 & 7.38 & 5.89 & 12.35 & 5.80 & 18.23 & 2.67 \\
\midrule

\multirow{7}{*}{\begin{tabular}[c]{@{}l@{}} \Design \\(5 stages) \end{tabular}}
&\cellcolor{gray!15}Multiplication &\cellcolor{gray!15}7.12 &\cellcolor{gray!15}5.55 &\cellcolor{gray!15}3.97 &\cellcolor{gray!15}3.25 &\cellcolor{gray!15}2.45 &\cellcolor{gray!15}3.10 &\cellcolor{gray!15}1.76 \\
& Scaled Addition & 9.16 & 6.22 & 7.40 & 3.13 &	2.65 & 2.56 & 1.46 \\
&\cellcolor{gray!15}Abs. Subtraction &\cellcolor{gray!15}12.96 &\cellcolor{gray!15}8.79 &\cellcolor{gray!15}14.68 &\cellcolor{gray!15}4.66 &\cellcolor{gray!15}4.00 &\cellcolor{gray!15}4.69 &\cellcolor{gray!15}2.28 \\
& Division & 16.50 & 12.14 & 21.52 & 7.37 &	5.98 & 7.54 & 4.93 \\
&\cellcolor{gray!15}Minimum&\cellcolor{gray!15}7.67 &\cellcolor{gray!15}5.60 &\cellcolor{gray!15}10.43 &\cellcolor{gray!15}3.26 &\cellcolor{gray!15}2.55 &\cellcolor{gray!15}2.70 &\cellcolor{gray!15}1.69 \\
& Maximum & 12.29 & 7.52 & 9.46 & 3.33 & 3.17 & 3.49 & 1.52 \\
\midrule

\multirow{7}{*}{\begin{tabular}[c]{@{}l@{}} \Design \\(6 stages) \end{tabular}}
&\cellcolor{gray!15}Multiplication &\cellcolor{gray!15}7.16 &\cellcolor{gray!15}5.54 &\cellcolor{gray!15}3.89 &\cellcolor{gray!15}2.55 &\cellcolor{gray!15}2.07 &\cellcolor{gray!15}2.13 &\cellcolor{gray!15}1.30 \\
& Scaled Addition & 9.25 & 7.48 & 4.18 & 3.03 &	2.18 & 1.90 & 1.20 \\
&\cellcolor{gray!15}Abs. Subtraction &\cellcolor{gray!15}14.11 &\cellcolor{gray!15}11.15 &\cellcolor{gray!15}5.77 &\cellcolor{gray!15}4.14 &\cellcolor{gray!15}3.07 &\cellcolor{gray!15}2.35 &\cellcolor{gray!15}1.80 \\
& Division & 18.70 & 13.14 & 8.44 & 6.32 & 5.03 & 4.41 & 3.79 \\
&\cellcolor{gray!15}Minimum&\cellcolor{gray!15}8.14 &\cellcolor{gray!15}5.72 &\cellcolor{gray!15}3.86 &\cellcolor{gray!15}2.87 &\cellcolor{gray!15}2.11 &\cellcolor{gray!15}1.65 &\cellcolor{gray!15}1.19 \\
& Maximum & 13.00 & 10.86 & 4.60 & 3.05 & 2.22 & 1.67 & 1.33 \\
\midrule

\multirow{7}{*}{\begin{tabular}[c]{@{}l@{}} \Design \\(7 stages) \end{tabular}}
&\cellcolor{gray!15}Multiplication &\cellcolor{gray!15}7.05 &\cellcolor{gray!15}4.95 &\cellcolor{gray!15}3.51 &\cellcolor{gray!15}2.53 &\cellcolor{gray!15}1.76 &\cellcolor{gray!15}2.32 &\cellcolor{gray!15}1.77 \\
& Scaled Addition & 8.85 & 6.00 & 4.77 & 9.78 &	2.18 & 1.48 & 1.68 \\
&\cellcolor{gray!15}Abs. Subtraction &\cellcolor{gray!15}12.15 &\cellcolor{gray!15}8.45 &\cellcolor{gray!15}8.63 &\cellcolor{gray!15}14.06 &\cellcolor{gray!15}3.16 &\cellcolor{gray!15}2.10 &\cellcolor{gray!15}2.85 \\
& Division & 15.96 & 11.75 & 14.05 & 13.05 & 5.03 &	4.12 & 6.59 \\
&\cellcolor{gray!15}Minimum&\cellcolor{gray!15}7.55 &\cellcolor{gray!15}5.50 &\cellcolor{gray!15}6.34 &\cellcolor{gray!15}5.24 &\cellcolor{gray!15}2.00 &\cellcolor{gray!15}1.42 &\cellcolor{gray!15}2.45 \\
& Maximum & 11.50 & 7.06 & 5.43 & 16.01 & 2.52 & 1.51 & 1.28 \\
\midrule

\multirow{7}{*}{\begin{tabular}[c]{@{}l@{}} \Design \\(8 stages) \end{tabular}}
&\cellcolor{gray!15}Multiplication &\cellcolor{gray!15}6.99 &\cellcolor{gray!15}5.19 &\cellcolor{gray!15}3.51 &\cellcolor{gray!15}2.50 &\cellcolor{gray!15}1.72 &\cellcolor{gray!15}1.37 &\cellcolor{gray!15}0.91 \\
& Scaled Addition & 8.93 & 6.00 & 4.65 & 2.91 &	2.05 & 1.47 & 1.04 \\
&\cellcolor{gray!15}Abs. Subtraction &\cellcolor{gray!15}12.41 &\cellcolor{gray!15}8.38 &\cellcolor{gray!15}8.20 &\cellcolor{gray!15}4.06 &\cellcolor{gray!15}2.88 &\cellcolor{gray!15}2.02 &\cellcolor{gray!15}1.49 \\
& Division & 16.15 & 11.72 & 14.17 & 6.47 &	5.00 & 4.08 & 3.66 \\
&\cellcolor{gray!15}Minimum&\cellcolor{gray!15}7.59 &\cellcolor{gray!15}5.45 &\cellcolor{gray!15}6.46 &\cellcolor{gray!15}2.78 &\cellcolor{gray!15}1.98 &\cellcolor{gray!15}1.41 &\cellcolor{gray!15}1.05 \\
& Maximum & 11.58 & 7.02 & 4.58 & 3.07 & 2.05 & 1.43 & 1.03 \\
\midrule
						
\multirow{7}{*}{\begin{tabular}[c]{@{}l@{}} \Design \\(16 stages) \end{tabular}}
&\cellcolor{gray!15}Multiplication &\cellcolor{gray!15}6.90 &\cellcolor{gray!15}4.85 &\cellcolor{gray!15}3.47 &\cellcolor{gray!15}2.48 &\cellcolor{gray!15}1.75 &\cellcolor{gray!15}1.26 &\cellcolor{gray!15}0.89 \\
& Scaled Addition & 8.80 & 5.97 & 4.15 & 2.89 & 2.04 & 1.43 & 1.03 \\
&\cellcolor{gray!15}Abs. Subtraction &\cellcolor{gray!15}12.23 &\cellcolor{gray!15}8.34 &\cellcolor{gray!15}5.82 &\cellcolor{gray!15}4.02 &\cellcolor{gray!15}2.82 &\cellcolor{gray!15}2.01 &\cellcolor{gray!15}1.47 \\
& Division & 16.20 & 11.69 & 8.62 & 6.24 & 4.93 & 4.01 & 3.54 \\
&\cellcolor{gray!15}Minimum&\cellcolor{gray!15}7.57 &\cellcolor{gray!15}5.44 &\cellcolor{gray!15}3.92 &\cellcolor{gray!15}2.75 &\cellcolor{gray!15}1.98 &\cellcolor{gray!15}1.39 &\cellcolor{gray!15}1.01 \\
& Maximum & 11.56 & 7.03 & 4.50 & 3.01 & 2.04 & 1.43 & 1.03 \\
\midrule

\multirow{7}{*}{\begin{tabular}[c]{@{}l@{}} \Design \\(32 stages) \end{tabular}}
&\cellcolor{gray!15}Multiplication &\cellcolor{gray!15}6.99 &\cellcolor{gray!15}4.95 &\cellcolor{gray!15}3.47 &\cellcolor{gray!15}2.45 &\cellcolor{gray!15}1.77 &\cellcolor{gray!15}1.26 &\cellcolor{gray!15}0.89 \\
& Scaled Addition & 8.84 & 5.98 & 4.15 & 2.88 &	2.01 & 1.43 & 1.04 \\
&\cellcolor{gray!15}Abs. Subtraction &\cellcolor{gray!15}12.28 &\cellcolor{gray!15}8.42 &\cellcolor{gray!15}5.76 &\cellcolor{gray!15}3.97 &\cellcolor{gray!15}2.81 &\cellcolor{gray!15}2.03 &\cellcolor{gray!15}1.46 \\
& Division & 16.00 & 11.70 & 8.38 &	6.32 & 4.91 & 4.03 & 3.50 \\
&\cellcolor{gray!15}Minimum&\cellcolor{gray!15}7.54 &\cellcolor{gray!15}5.45 &\cellcolor{gray!15}3.85 &\cellcolor{gray!15}2.71 &\cellcolor{gray!15}1.95 &\cellcolor{gray!15}1.41 &\cellcolor{gray!15}1.02 \\
& Maximum & 11.56 & 7.02 & 4.49 & 3.00 & 2.01 & 1.43 & 1.02 \\
\midrule

\multirow{7}{*}{\begin{tabular}[c]{@{}l@{}} \Design \\(64 stages) \end{tabular}}
&\cellcolor{gray!15}Multiplication &\cellcolor{gray!15}6.85 &\cellcolor{gray!15}5.04 &\cellcolor{gray!15}3.45 &\cellcolor{gray!15}2.45 &\cellcolor{gray!15}1.76 &\cellcolor{gray!15}1.23 &\cellcolor{gray!15}0.89 \\
& Scaled Addition & 8.86 & 6.02 & 4.10 & 2.88 &	2.02 & 1.45 & 1.04 \\
&\cellcolor{gray!15}Abs. Subtraction &\cellcolor{gray!15}12.40 &\cellcolor{gray!15}8.42 &\cellcolor{gray!15}5.79 &\cellcolor{gray!15}4.01 &\cellcolor{gray!15}2.84 &\cellcolor{gray!15}2.05 &\cellcolor{gray!15}1.45 \\
& Division & 16.14 & 11.61 & 8.54 & 6.21 & 4.89	& 4.03 & 3.54 \\
&\cellcolor{gray!15}Minimum&\cellcolor{gray!15}7.67 &\cellcolor{gray!15}5.42 &\cellcolor{gray!15}3.85 &\cellcolor{gray!15}2.72 &\cellcolor{gray!15}1.93 &\cellcolor{gray!15}1.40 &\cellcolor{gray!15}1.00 \\
& Maximum & 11.53 & 7.08 & 4.48 & 2.99 & 2.03 & 1.45 & 1.02 \\
\midrule

\multirow{7}{*}{\begin{tabular}[c]{@{}l@{}} \Design \\(128 stages) \end{tabular}}
&\cellcolor{gray!15}Multiplication &\cellcolor{gray!15}6.95 &\cellcolor{gray!15}4.89 &\cellcolor{gray!15}3.51 &\cellcolor{gray!15}2.44 &\cellcolor{gray!15}1.74 &\cellcolor{gray!15}1.25 &\cellcolor{gray!15}0.90 \\
& Scaled Addition & 8.85 & 6.01 & 4.13 & 2.88 &	2.03 & 1.44 & 1.04 \\
&\cellcolor{gray!15}Abs. Subtraction &\cellcolor{gray!15}12.36 &\cellcolor{gray!15}8.24 &\cellcolor{gray!15}5.70 &\cellcolor{gray!15}3.97 &\cellcolor{gray!15}2.81 &\cellcolor{gray!15}2.01 &\cellcolor{gray!15}1.48 \\
& Division & 16.16 & 11.41 & 8.33 &	6.09 & 4.85 & 4.04 & 3.56 \\
&\cellcolor{gray!15}Minimum&\cellcolor{gray!15}7.69 &\cellcolor{gray!15}5.40 &\cellcolor{gray!15}3.84 &\cellcolor{gray!15}2.75 &\cellcolor{gray!15}1.94 &\cellcolor{gray!15}1.39 &\cellcolor{gray!15}1.02 \\
& Maximum & 11.59 & 7.06 & 4.47 & 2.96 & 2.03 & 1.41 & 1.03 \\
\midrule

% Data Section 7
\multirow{7}{*}{\begin{tabular}[c]{@{}l@{}} \texttt{MATLAB} \\ (rand)\end{tabular}}
&\cellcolor{gray!15}Multiplication &\cellcolor{gray!15}6.97 &\cellcolor{gray!15}4.90 &\cellcolor{gray!15}3.50 &\cellcolor{gray!15}2.43 &\cellcolor{gray!15}1.72 &\cellcolor{gray!15}1.23 &\cellcolor{gray!15}0.86 \\
& Scaled Addition & 9.05 & 6.38 & 4.51 & 3.18 & 2.23 & 1.60 & 1.13 \\
%& Approx Add & 14.8 & 11.5 & 9.19 & 8.05 & 7.25 & 6.83 & 6.48 & 6.40 \\
&\cellcolor{gray!15}Abs. Subtraction &\cellcolor{gray!15}7.75 &\cellcolor{gray!15}5.47 &\cellcolor{gray!15}3.92 &\cellcolor{gray!15}2.77 &\cellcolor{gray!15}1.97 &\cellcolor{gray!15}1.38 &\cellcolor{gray!15}0.98 \\
& Division & 12.1 & 8.59 & 6.17 & 4.37 & 3.12 & 2.17 & 1.55 \\
&\cellcolor{gray!15}Minimum &\cellcolor{gray!15}7.89 &\cellcolor{gray!15}5.54 &\cellcolor{gray!15}3.92 &\cellcolor{gray!15}2.78 &\cellcolor{gray!15}1.94 &\cellcolor{gray!15}1.37 &\cellcolor{gray!15}0.99 \\
& Maximum & 7.94 & 5.52 & 3.90 & 2.74 & 1.96 & 1.38 & 0.98 \\
\midrule

% Data Section 8
\multirow{7}{*}{\begin{tabular}[c]{@{}l@{}}PRNG\\ ($\log_2{L}$-bit LFSR) \end{tabular}}
&\cellcolor{gray!15}Multiplication &\cellcolor{gray!15}6.87 &\cellcolor{gray!15}4.09 &\cellcolor{gray!15}2.55 &\cellcolor{gray!15}1.62 &\cellcolor{gray!15}1.07 &\cellcolor{gray!15}0.70 &\cellcolor{gray!15}0.48 \\
& Scaled Addition & 8.54 & 5.32 & 3.18 & 1.97 & 1.32 & 0.82 & 0.26 \\
%& Approx. Add & 21.3 & 13.8 & 10.3 & 8.22 & 7.22 & 6.86 & 6.48 & 6.41 \\
&\cellcolor{gray!15}Abs. Subtraction &\cellcolor{gray!15}5.35 &\cellcolor{gray!15}2.88 &\cellcolor{gray!15}1.43 &\cellcolor{gray!15}0.68 &\cellcolor{gray!15}0.42 &\cellcolor{gray!15}0.20 &\cellcolor{gray!15}0.10 \\
& Division & 13.6 & 10.54 & 9.50 & 8.74 & 8.82 & 8.31 & 8.39 \\
&\cellcolor{gray!15}Minimum &\cellcolor{gray!15}6.92 &\cellcolor{gray!15}3.81 &\cellcolor{gray!15}1.96 &\cellcolor{gray!15}0.93 &\cellcolor{gray!15}0.57 &\cellcolor{gray!15}0.23 &\cellcolor{gray!15}0.15 \\
& Maximum & 4.96 & 2.64 & 1.38 & 0.68 & 0.37 & 0.14 & 0.11 \\
\midrule

% Data Section 9
\multirow{7}{*}{\begin{tabular}[c]{@{}l@{}}QRNG\\(Sobol)\end{tabular}}
&\cellcolor{gray!15}Multiplication &\cellcolor{gray!15}3.88 &\cellcolor{gray!15}1.98 &\cellcolor{gray!15}1.05 &\cellcolor{gray!15}0.58 &\cellcolor{gray!15}0.27 &\cellcolor{gray!15}0.14 &\cellcolor{gray!15}0.07 \\
& Scaled Addition & 3.79 & 2.59 & 0.94 & 0.47 & 0.33 & 0.12 & 0.06 \\
%& Approx. Add & 4.16 & 2.08 & 1.04 & 0.52 & 0.26 & 0.13 & 0.06 & 0.03 \\
&\cellcolor{gray!15}Abs. Subtraction &\cellcolor{gray!15}2.08 &\cellcolor{gray!15}1.04 &\cellcolor{gray!15}0.52 &\cellcolor{gray!15}0.26 &\cellcolor{gray!15}0.13 &\cellcolor{gray!15}0.06 &\cellcolor{gray!15}0.03 \\
& Division & 4.86 & 3.26 & 2.86 & 2.71 & 2.72 & 2.76 & 2.75 \\
&\cellcolor{gray!15}Minimum &\cellcolor{gray!15}3.19 &\cellcolor{gray!15}1.57 &\cellcolor{gray!15}0.79 &\cellcolor{gray!15}0.39 &\cellcolor{gray!15}0.20 &\cellcolor{gray!15}0.10 &\cellcolor{gray!15}0.05 \\
& Maximum & 3.09 & 1.56 & 0.77 & 0.39 & 0.20 & 0.10 & 0.05 \\
\midrule

\multirow{7}{*}{\begin{tabular}[c]{@{}l@{}}TRUE-BSG~\cite{TRUE_BSG_ISCAS2025}\end{tabular}}
&\cellcolor{gray!15}Multiplication &\cellcolor{gray!15}7.05 &\cellcolor{gray!15}5.02 &\cellcolor{gray!15}3.52 &\cellcolor{gray!15}2.47 &\cellcolor{gray!15}1.74 &\cellcolor{gray!15}1.24 &\cellcolor{gray!15}0.88 \\
& Scaled Addition & 8.98 & 6.42 & 4.63 & 3.20 & 2.30 & 1.66 & 1.14 \\
&\cellcolor{gray!15}Abs. Subtraction &\cellcolor{gray!15}7.82 &\cellcolor{gray!15}5.63 &\cellcolor{gray!15}3.95 &\cellcolor{gray!15}2.76 &\cellcolor{gray!15}1.92 &\cellcolor{gray!15}1.37 &\cellcolor{gray!15}0.96 \\
& Division & 12.14 & 8.87 & 6.18 & 4.30 & 3.02 & 2.16 & 1.50 \\
&\cellcolor{gray!15}Minimum &\cellcolor{gray!15}7.85 &\cellcolor{gray!15}5.74 &\cellcolor{gray!15}4.04 &\cellcolor{gray!15}2.82 &\cellcolor{gray!15}1.97 &\cellcolor{gray!15}1.41 &\cellcolor{gray!15}0.98 \\
& Maximum & 7.65 & 5.59 & 4.02 & 2.79 & 2.00 & 1.45 & 0.99 \\
\midrule

\multirow{7}{*}{\begin{tabular}[c]{@{}l@{}} IMSNG~\cite{SC_IMC_DAC2025}\end{tabular}}
&\cellcolor{gray!15}Multiplication &\cellcolor{gray!15}7.00 &\cellcolor{gray!15}5.03 &\cellcolor{gray!15}3.60 &\cellcolor{gray!15}2.63 &\cellcolor{gray!15}2.03 &\cellcolor{gray!15}1.67 &\cellcolor{gray!15}1.45 \\
& Scaled Addition & 9.05 & 6.42 & 4.57 & 3.37 & 2.46 & 1.89 & 1.52 \\
&\cellcolor{gray!15}Abs. Subtraction &\cellcolor{gray!15}8.16 &\cellcolor{gray!15}5.86 &\cellcolor{gray!15}4.35 &\cellcolor{gray!15}3.33 &\cellcolor{gray!15}2.67 &\cellcolor{gray!15}2.32 &\cellcolor{gray!15}2.20 \\
& Division & 12.14 & 8.74 & 6.51 & 4.78 & 3.63 & 2.93 & 2.67 \\
&\cellcolor{gray!15}Minimum &\cellcolor{gray!15}7.81 &\cellcolor{gray!15}5.55 &\cellcolor{gray!15}4.00 &\cellcolor{gray!15}2.98 &\cellcolor{gray!15}2.28 &\cellcolor{gray!15}1.83 &\cellcolor{gray!15}1.67 \\
& Maximum & 7.84 & 5.62 & 4.11 & 3.10 & 2.41 & 2.05 & 1.86 \\

\bottomrule
\end{tabular}
}
\label{tab:sc_arithmetic_comparison}
\vspace{-1em}
\end{table}

%\vspace{-0.5em}
\subsection{{Hardware costs comparison}}
%\subsubsection{\textbf{Hardware costs comparison}}
Next, we compare the hardware costs of the proposed generator and the SOTA LFSR- and Sobol-based bit-stream generation methods. 
We described all designs in Verilog HDL and synthesized them using the Synopsys Design Compiler v2024.09-SP4 with the 45$nm$ FreePDK gate library~\cite{FreePDK45}.
Table~\ref{hw_cost_RNG} details the hardware costs %comparison
of different %bit-stream generation
approaches. 

The \Design~exhibits linear hardware scaling. As the state space expands from 4 to 8 stages, the hardware footprint area increases proportionally from 54 to 116$\mu m^2$. This linear growth confirms the structural efficiency of the cascaded \texttt{MUX} chain. The critical path latency (CPL) also scales predictably with the addition of stages, reflecting the physical propagation delay of the signal race condition.
The %optimal 
\Design~16 stage configuration achieves silicon area of 240$\mu m^2$ close to the common LFSR--comparator baseline (233$\mu m^2$). However, traditional LFSRs provide only deterministic pseudo-randomness and suffer from autocorrelation. %in sequential SC logic, 
The \Design~designs deliver a reliable TRNG alongside intrinsic physical hardware unclonability with no %virtually zero 
area penalty.

When benchmarked against SOTA works, the \Design~8 stage demonstrates competitive area and speed advantages. To ensure a fair architectural comparison against TRUE-BSG~\cite{TRUE_BSG_ISCAS2025} (which relies on a high-speed Ring Oscillator), the cost of the required 8-bit binary comparator has been %explicitly 
integrated into the TRUE-BSG baseline, resulting in a total area of 295 $\mu m^2$ and an energy consumption of 0.05 pJ/cycle. Compared to this adjusted baseline, the proposed \Design~8 stage 
is 2.5$\times$ smaller
%achieves a significantly smaller silicon footprint 
(116 vs. 295 $\mu m^2$). 
While TRUE-BSG uses less energy per cycle, the \Design~8 design eliminates the need for a multi-bit comparator and provides the dual utility of SC bit-stream generation and device authentication without requiring a dedicated cryptographic core.
Furthermore, the \Design~8 stage outperforms alternative-domain generators like the IMSNG design of~\cite{SC_IMC_DAC2025}. Operating entirely within ReRAM arrays, IMSNG suffers from a high conversion latency (78.2 $ns$) and significant energy consumption (3,420 $pJ$/cycle) due to the overhead of heavy mixed-signal read operations, reference current generation, and peripheral circuitry. In contrast, the \Design~8 design generates SC bit-streams over two orders of magnitude faster (CPL of 0.41 $ns$) at 0.34 $pJ$ energy per cycle.
While previous accuracy evaluations demonstrated that quasi-random Sobol sequences offer the lowest theoretical MAE, Table~\ref{hw_cost_RNG} demonstrates their high hardware cost.  The Sobol generator with comparator occupies 850 $\mu m^2$, 7$\times$ the \Design~8 stage. Even when optimized into an FSM-based topology (281 $\mu m^2$), the Sobol generator remains physically larger than the \Design~8 stage and lacks any physical unclonability features.

%==================================================================
%       Mehran - Accuracy Table for HW measurements
%==================================================================
\begin{table}[!t]
\centering
\caption{Hardware Cost of Diverse Methods for a \textit{\underline{Single}} SC Bit-stream Generation. (The designs are synthesized in 45$nm$ for 8-bit precision.)}
%\begin{threeparttable}
\vspace{-0.75em}
\renewcommand{\arraystretch}{0.8}
\resizebox{\columnwidth}{!}{
\begin{tabular}{lccccc} 
\toprule
%\textbf{SBS Generation Method} 
\begin{tabular}[c]{@{}c@{}} \textbf{Method}\end{tabular}
& \begin{tabular}[c]{@{}c@{}}\textbf{Area} \\($\mu m^2$)\end{tabular} &  \begin{tabular}[c]{@{}c@{}}\textbf{CPL} \\($ns$)\end{tabular} & \begin{tabular}[c]{@{}c@{}}\textbf{Power@}\\\textbf{Max Freq.} \\($m W$)\end{tabular} & \begin{tabular}[c]{@{}c@{}}\textbf{Energy}\\\textbf{/Cycle} \\($pJ$)\end{tabular} & \begin{tabular}[c]{@{}c@{}}\textbf{Area$\times$Delay} \\($\mu m^2\times ns$)\end{tabular} \\ 
\midrule
\Design--4 stages & 54 & 0.28 & 0.59 & 0.16 & 15.1\\
\Design--5 stages & 70 & 0.31 & 0.63 & 0.19 & 21.7\\
\Design--6 stages & 85 & 0.34 & 0.75 & 0.25 & 28.9\\
\Design--7 stages & 101 & 0.45 & 0.65 & 0.29 & 45.4\\
\Design--8 stages & 116 & 0.41 & 0.84 & 0.34 & 47.6\\
\Design--16 stages & 240 & 0.67 & 1.04 & 0.70 & 160.8\\
\midrule
LFSR + Comparator & 233 & 0.37 & 0.95 & 0.35 & 86.2\\ 
Sobol + Comparator~\cite{liu2017energy} & 850 & 0.45 & 1.31 & 0.59 & 382\\ 
Sobol (FSM-based)~\cite{Najafi_FSM_DATE2021} & 281 & 0.36 & 0.72 & 0.26 & 101\\ 
TRUE-BSG~\cite{TRUE_BSG_ISCAS2025} & 295 & -- & -- & 0.05 & --\\
IMSNG~\cite{SC_IMC_DAC2025} & -- & 78.2 & -- & 3420 & --\\

\bottomrule
\end{tabular}
}
%\vspace{-1em}
\parbox{\columnwidth}{\scriptsize 
}
\label{hw_cost_RNG}
\vspace{-1.0em}
\end{table}

%\noindent \hl{\textbf{\textit{Corollary:}}} If a target application can tolerate a slight relaxation in computational MAE, the 8-stage \Design~presents an optimum solution. At only 116 $\mu m^2$ and 0.34 $pJ$ per cycle, it completely breaks the hardware floor of traditional LFSRs, reducing area by 50\% while still preserving physical unclonability. This flexibility allows designers to precisely tune the core SC  computation engine for either peak accuracy ($N=16$) or extreme area constraint ($N=8$) using the exact same underlying logic primitive.
%\vspace{-0.5em}
\subsection{Overall Design Efficiency }
%\subsubsection{Performance Analysis}
To further validate the effectiveness of the  
\Design~design, we evaluate the \textit{Figure of Merit (FoM)}, a metric that combines accuracy and hardware (HW) costs to reflect overall design efficiency: %\textit{FoM} is defined as:
\textit{FoM} = $\frac{\text{Accuracy}}{\text{HW Costs}} = \frac{1}{(\text{MAE}\times\text{Area}\times\text{Power}\times\text{Latency})}$.
%\vspace{-0.5em}
%\begin{equation}
%    \label{FoM_eq}
%    FoM = \frac{Accuracy}{HW Costs}
%        = \frac{\frac{1}{MAE}}{Area\times Power\times Latency}
%\end{equation}
%\vspace{-.75em}
\noindent %This metric provides a unified basis for comparing different designs by jointly considering MAE, hardware footprint, power consumption, and latency.
Fig.~\ref{fig:fom} presents the \textit{FoM} evaluation across various designs, using data extracted from Tables~\ref{tab:accuracy_bs}, \ref{tab:sc_arithmetic_comparison}, and \ref{hw_cost_RNG}. %A higher \textit{FoM} value indicates superior overall performance, achieved through a combination of higher computational accuracy and reduced hardware cost.
The results consistently show that the 
\Design--8 design achieves a comparable \textit{FoM} to TRUE-BSG~\cite{TRUE_BSG_ISCAS2025} %as a hardware-based true-random design
across all SC bit-stream generation and arithmetic operations. %the highest FoM among all hardware true-random designs across all evaluated SC operations, outperforming both TRUE-BSG and IMSNG by a significant margin.
While the LFSR- and Sobol-based designs achieve lower MAE, their \textit{FoM} surpasses all the other designs for SC arithmetic operations.
%While the LFSR- and Sobol-based designs achieve lower MAE, their substantially higher hardware cost results in a considerably lower FoM, confirming that their accuracy advantage does not translate into overall design efficiency. 
Among all \Design~configurations, the 8-stage design consistently occupies the 
optimal operating point, offering the best balance between 
accuracy and hardware cost. Deeper configurations (16 stages 
and beyond) provide no significant accuracy benefit while incurring higher area and power, driving their \textit{FoM} down. %These results validate the \Design--8 stage as the most efficient overall design for true-random stochastic bit-stream generation.

\vspace{-0.25em} 
\subsection{Calibration, Storage, and Stability}
\label{sec:stability}
Calibration happens once per chip, offline. The $N+1$ entries of $\vec{w}$ are fitted from a few hundred challenge-response pairs. Any PUF deployment already collects these during enrollment, so nothing new is measured. Building the table for all 255 8-bit levels then takes 43\,ms for $N=16$ in single-threaded Python on an Intel Core i7-12700H, and the $1.1\times10^6$ multiply-accumulates it requires would take well under a millisecond in a small on-chip calibration engine. Each level stores one $N$-bit challenge, which is 256\,B for $N=8$ and 512\,B for $N=16$. An application that only uses a fixed set of constants stores only those.
The table also holds up under operating conditions. A voltage or temperature shift slows both paths of the race by nearly the same amount, so it drops out of $\Delta$. The noise $\sigma_{dynamic}$ scales with the same nominal delay, and $p$ depends only on the ratio $\frac{\Delta}{\sigma_{dynamic}}$. What remains is second order. In the model, a 20\% change in noise scale adds 2.9\% to 3.5\% MAE, and a fixed offset of $0.1\,\sigma_{dynamic}$, the kind aging or arbiter mismatch would introduce, adds 2.8\%. Neither produces a cliff. Levels near $p=0.5$ are the least affected because noise dominates there. If drift ever exceeds this, recalibration is the same sub-second offline step and can be triggered by an on-die sensor.

%===========================================
%       New Plot for FoM
%===========================================
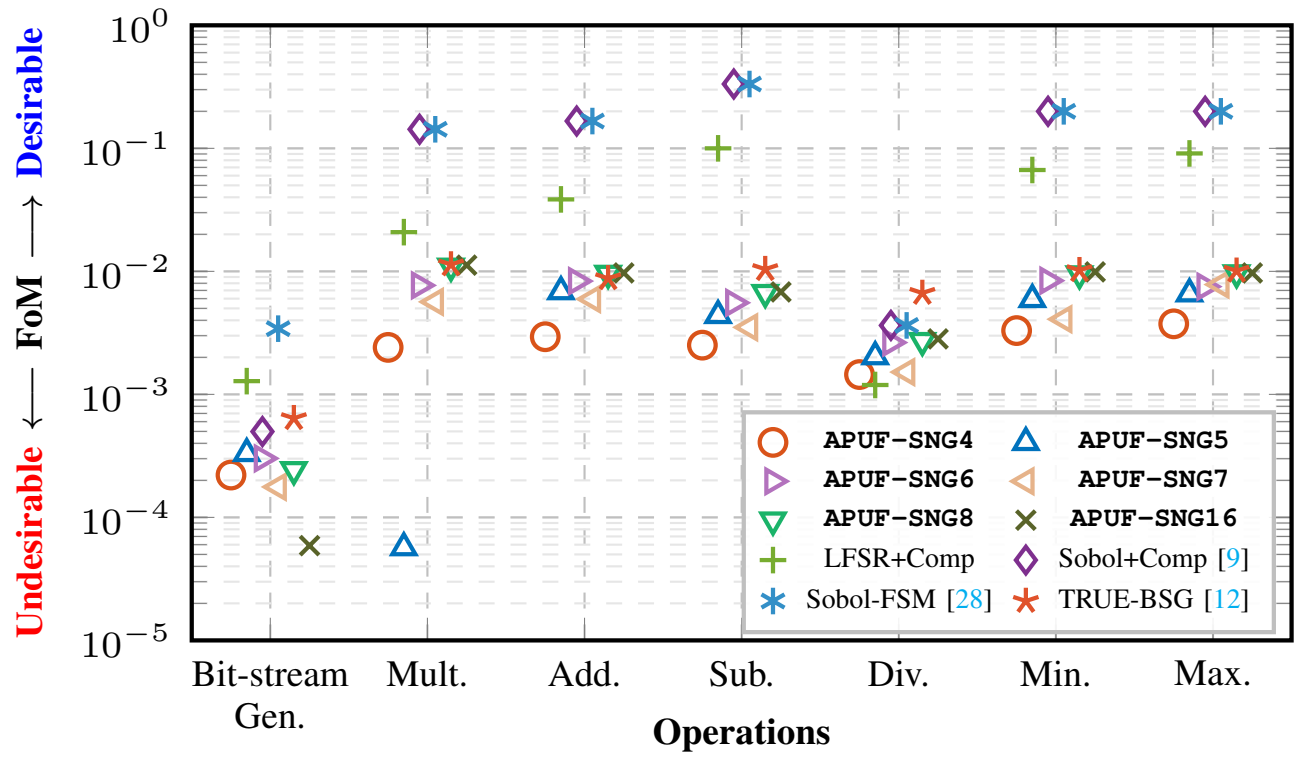
\begin{figure}[!t]
%\vspace{-0.5em} 
    \centering
    \begin{tikzpicture}[scale=1]
\begin{axis}[
    width=\columnwidth, 
    height=5.65cm,       
    font=\sffamily, 
    ymode=log,
    ymin=1e-5, ymax=1.0e0,
    xmin=0.5, xmax=7.5,
    grid=both,
    grid style={dashed, gray!50},
    minor grid style={dashed, gray!20},
    xtick={1,2,3,4,5,6,7},
    xticklabels={Bit-stream\\Gen., Mult., Add., Sub., Div., Min., Max.},
    xticklabel style={align=center, font=\scriptsize},
    xlabel={\textbf{Operations}},
    xlabel style={font=\scriptsize, yshift=2ex},
    yticklabel style={font=\scriptsize},
    ylabel={\textcolor{red}{\textbf{Undesirable}} $\longleftarrow$ \textbf{FoM} $\longrightarrow$ \textcolor{blue}{\textbf{Desirable}}},
    ylabel style={font=\scriptsize\bfseries, align=center},
    legend columns=2, 
    legend style={
        at={(0.5,0.01)}, 
        anchor=south west,
        font=\scriptsize, 
        cells={align=left}, 
        nodes={scale=0.75, transform shape},
        column sep=0.5ex,
        draw=gray!50,
        fill=white,
        fill opacity=0.9,
        text opacity=1
    },
    only marks,
    mark size=2.5pt, 
    thick
]

% APUF-SNG4
\addplot [color=cAPUF4, mark=o] coordinates {
    (0.75, 2.2088e-4) (1.75, 2.4096e-3) (2.75, 2.9411e-3) (3.75, 2.5125e-3) 
    (4.75, 1.4492e-3) (5.75, 3.3003e-3) (6.75, 3.7453e-3)
};
\addlegendentry{\textbf{\texttt{APUF-SNG4}}}

% APUF-SNG5
\addplot [color=cAPUF5, mark=triangle] coordinates {
    (0.85, 3.3122e-4) (1.85, 5.6818e-5) (2.85, 6.8493e-3) (3.85, 4.3859e-3) 
    (4.85, 2.0283e-3) (5.85, 5.9171e-3) (6.85, 6.5789e-3)
};
\addlegendentry{\textbf{\texttt{APUF-SNG5}}}

% APUF-SNG6
\addplot [color=cAPUF6, mark=triangle, every mark/.append style={rotate=-90}] coordinates {
    (0.95, 3.0165e-4) (1.95, 7.6923e-3) (2.95, 8.3333e-3) (3.95, 5.5555e-3) 
    (4.95, 2.6385e-3) (5.95, 8.4033e-3) (6.95, 7.5187e-3)
};
\addlegendentry{\textbf{\texttt{APUF-SNG6}}}

% APUF-SNG7
\addplot [color=cAPUF7, mark=triangle, every mark/.append style={rotate=90}] coordinates {
    (1.05, 1.7689e-4) (2.05, 5.6497e-3) (3.05, 5.9523e-3) (4.05, 3.5087e-3) 
    (5.05, 1.5174e-3) (6.05, 4.0816e-3) (7.05, 7.8125e-3)
};
\addlegendentry{\textbf{\texttt{APUF-SNG7}}}

% APUF-SNG8
\addplot [color=cAPUF8, mark=triangle, every mark/.append style={rotate=180}] coordinates {
    (1.15, 2.4379e-4) (2.15, 1.0989e-2) (3.15, 9.6153e-3) (4.15, 6.7114e-3) 
    (5.15, 2.7322e-3) (6.15, 9.5238e-3) (7.15, 9.7087e-3)
};
\addlegendentry{\textbf{\texttt{APUF-SNG8}}}

% APUF-SNG16
\addplot [color=cAPUF16, mark=x] coordinates {
    (1.25, 5.8934e-5) (2.25, 1.1235e-2) (3.25, 9.7087e-3) (4.25, 6.8027e-3) 
    (5.25, 2.8248e-3) (6.25, 9.9009e-3) (7.25, 9.7087e-3)
};
\addlegendentry{\textbf{\texttt{APUF-SNG16}}}

% LFSR+Comp
\addplot [color=cLFSR, mark=+] coordinates {
    (0.85, 1.2812e-3) (1.85, 2.0833e-2) (2.85, 3.8461e-2) (3.85, 1.0000e-1) 
    (4.85, 1.1918e-3) (5.85, 6.6666e-2) (6.85, 9.0909e-2)
};
\addlegendentry{LFSR+Comp}

% Sobol+Comp
\addplot [color=cSobolC, mark=diamond] coordinates {
    (0.95, 4.9850e-4) (1.95, 1.4285e-1) (2.95, 1.6666e-1) (3.95, 3.3333e-1) 
    (4.95, 3.6363e-3) (5.95, 2.0000e-1) (6.95, 2.0000e-1)
};
\addlegendentry{Sobol+Comp~\cite{liu2017energy}}

% Sobol-FSM
\addplot [color=cSobolF, mark=asterisk] coordinates {
    (1.05, 3.4218e-3) (2.05, 1.4285e-1) (3.05, 1.6666e-1) (4.05, 3.3333e-1) 
    (5.05, 3.6363e-3) (6.05, 2.0000e-1) (7.05, 2.0000e-1)
};
\addlegendentry{Sobol-FSM~\cite{Najafi_FSM_DATE2021}}

% TRUE-BSG
\addplot [color=cTRUE, mark=star] coordinates {
    (1.15, 6.3959e-4) (2.15, 1.1363e-2) (3.15, 8.7719e-3) (4.15, 1.0416e-2) 
    (5.15, 6.6666e-3) (6.15, 1.0204e-2) (7.15, 1.0101e-2)
};
\addlegendentry{TRUE-BSG~\cite{TRUE_BSG_ISCAS2025}}

\end{axis}
\end{tikzpicture}
\vspace{-1em} 

\caption{Comparison of \textit{FoM} across different design approaches.}
\label{fig:fom}
\vspace{-1.0em}
\end{figure}

%\section{Limitations and Future Work}
%Some practical concerns remain.
%First, decorrelation between generated stochastic streams must be verified carefully, because correlated streams break the simple probability multiplication rule. 
%Second, 
%Supply and temperature variation can shift the effective delay boundary, changing \(P(S=1)\) for a fixed challenge. Second, the challenge search used here is based on random sampling; more advanced strategies such as guided optimization, importance sampling, or machine-learning-assisted search can improve coverage of the near-boundary space. %Third, a full circuit-level study in Cadence should include parasitics, metastability characterization, Monte Carlo sweeps, and transient-noise-aware evaluation of the StrongARM arbiter.
%\vspace{-0.25em}
\section{Conclusion}
%\vspace{-0.1em}
This work repurposed the arbiter PUF as a stochastic number generator. Operating the APUF near its metastable point turns the path delay difference into a controllable output probability, so a stored challenge replaces the random number generator and comparator of a conventional SNG. A delay-and-noise model maps each target probability to a challenge, and enumeration of the $2^N$ challenge space finds it. Simulation shows that an 8-stage \Design{} reaches 1.04\% MAE at 1024 bits, matching TRUE-BSG and a software TRNG, with bit-streams that are uncorrelated ($\text{SCC}\, \approx\, 0$). Synthesized in 45\,$nm$, the 8-stage design occupies 116\,$\mu$m$^2$ and consumes 0.34\,pJ per cycle, 2.5$\times$ smaller than TRUE-BSG and 7$\times$ smaller than a Sobol generator with comparator.

\bibliographystyle{IEEEtran}
\bibliography{References/references}

%\newpage
%\vfill

\begin{IEEEbiography}[{\includegraphics[width=1.0in,height=1.0in,clip,keepaspectratio]{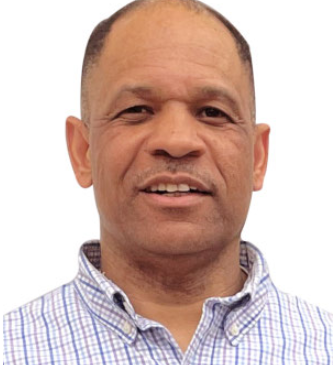}}]{Shelby Williams} (Member, IEEE) received the
B.Sc. and M.Sc. degrees in electrical engineering and computer science from the University of Louisiana at Lafayette, Lafayette, LA, USA, in 1996
and 2020, respectively, and the Ph.D. degree in computer engineering from the University of Louisiana at Lafayette. His current research interests include electronics, VLSI, microelectronics, random number
generators, and spiking neurons.
\end{IEEEbiography}

\begin{IEEEbiography}[{\includegraphics[width=1.3in,height=1.25in,clip,keepaspectratio]{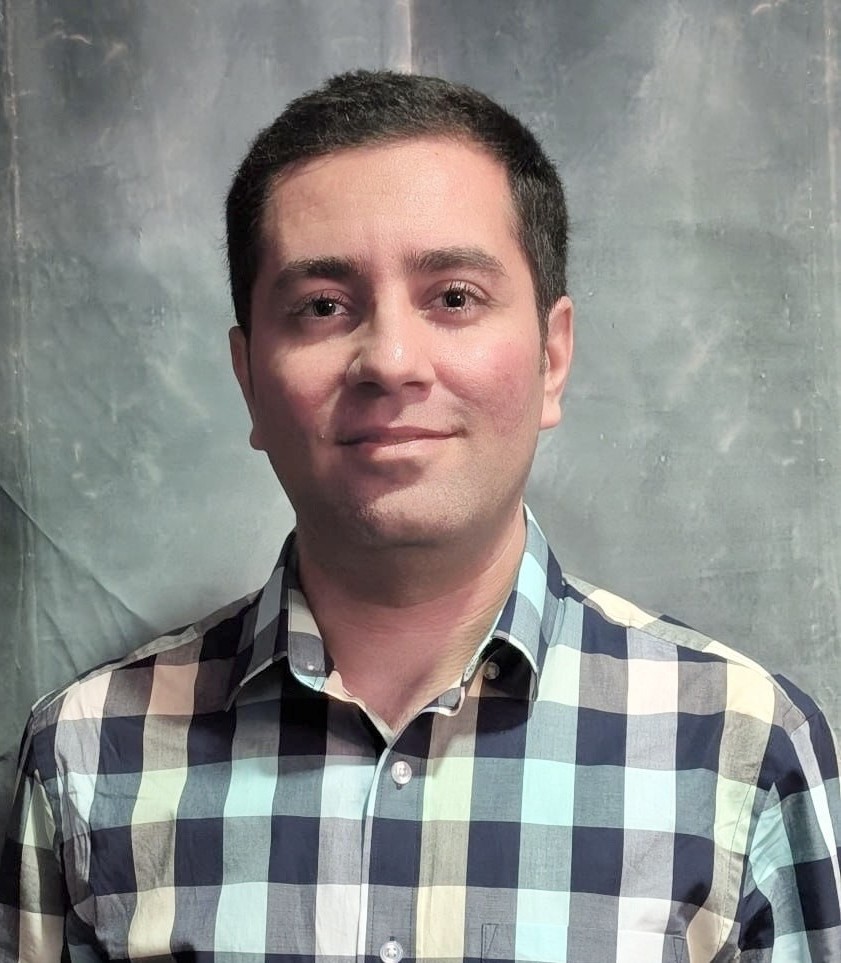}}]{Mehran Moghadam} (S’22) received the B.Sc. degree in Computer Engineering, and M.Sc. degree in \textit{Computer Systems Architecture} from the University of Isfahan, Iran, in 2010 and 2016. He graduated as one of the top-ranking students in both programs.
In 2022, he began his Ph.D. studies in Computer Engineering at the School of Computing and Informatics, University of Louisiana at Lafayette, Lafayette, LA, USA. In 2024, he transferred to the Department of Electrical, Computer, and Systems Engineering at Case Western Reserve University, Cleveland, OH, USA, to continue pursuing his Ph.D. in Computer Engineering.
He became a finalist in the ACM SIGBED Student Research Competition (SRC) at ESWEEK and ICCAD in 2024 and was selected as a Young Fellow in DAC 2024 and 2026. %He has authored/co-authored more than 30 peer-reviewed papers in top EDA venues.
Mehran received the Best Paper Award at ISLPED'26 conference. Furthermore, he received the IEEE Circuits and Systems Society (CASS) Pre-Doctoral Grant Award in 2026.
His research interests include emerging and unconventional computing paradigms, such as energy-efficient stochastic computing models, real-time and highly-accurate hyperdimensional computing systems, bit-stream processing, robust in-memory arithmetic computation, and low-power in/near-sensor computing designs for edge AI. 
\end{IEEEbiography}

\begin{IEEEbiography}[{\includegraphics[width=1.3in,height=1.25in,clip,keepaspectratio]{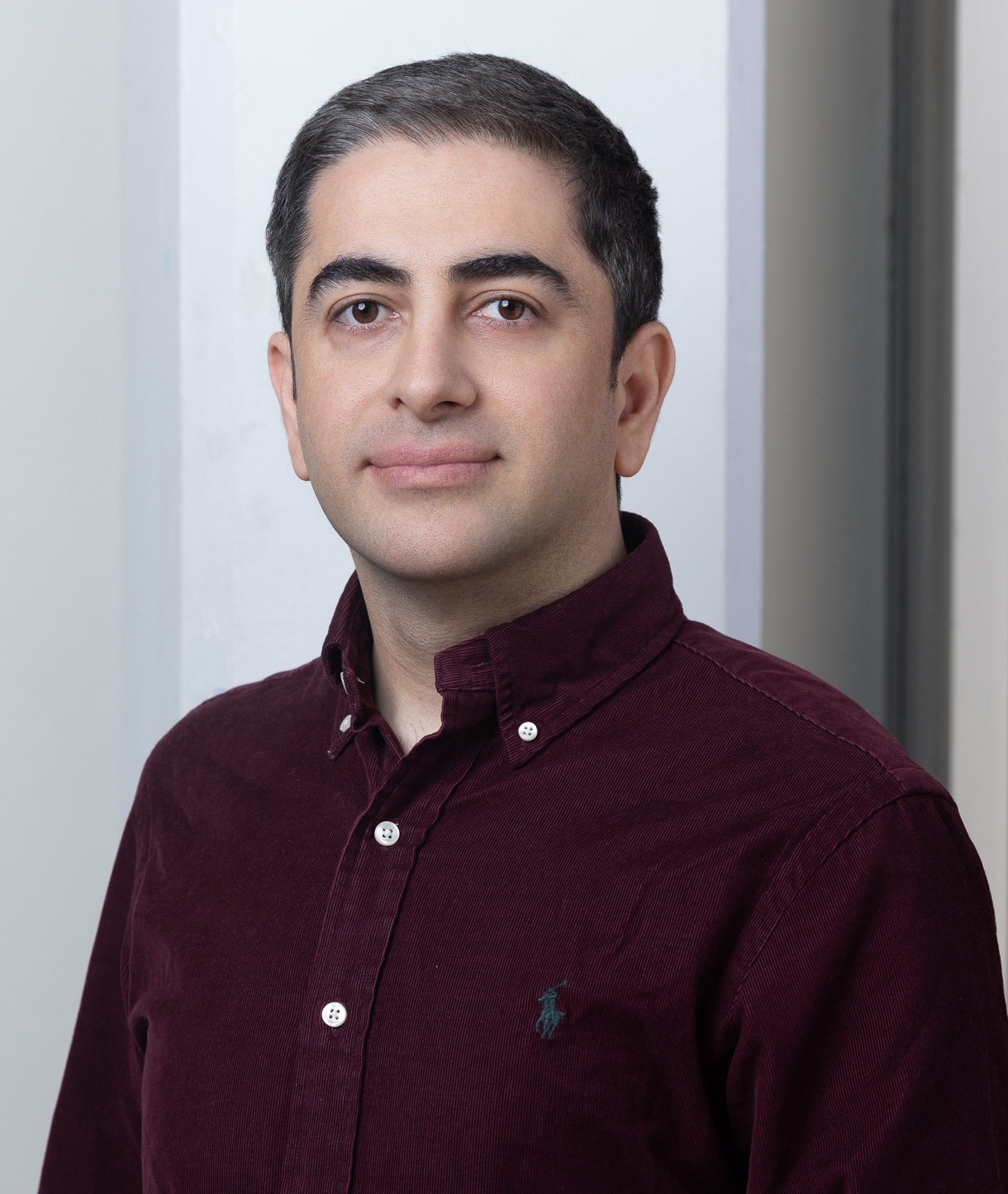}}]{M. Hassan Najafi} (S’15-M’18-SM'23) received the B.Sc. degree in Computer Engineering from the University of Isfahan, Iran, the M.Sc. degree in Computer Architecture from the University of Tehran, Iran, and the Ph.D. degree in Electrical Engineering from University of Minnesota, Twin Cities, USA, in 2011, 2014, and 2018, respectively. He was an Assistant Professor at the School of Computing and Informatics, University of Louisiana at Lafayette, Lafayette, LA, USA, from 2018 to 2024. He is currently an Associate Professor at the Electrical, Computer, and Systems Engineering Department at Case Western Reserve University, Cleveland, OH, USA. His research interests include stochastic and approximate computing, unary processing, in-memory computing, and hyperdimensional computing. He has authored/co-authored more than 120 peer-reviewed papers and has been granted 12 U.S. patents, with more pending. Dr. Najafi received the NSF CAREER Award in 2024, the Best Paper Award at GLSVLSI'23 and ICCD’17, the Best Poster Awards at GLSVLSI'24 and DCAS'26, the 2018 EDAA Outstanding Dissertation Award, and the Doctoral Dissertation Fellowship from the University of Minnesota. He has been an editor for the IEEE Journal on Emerging and Selected Topics in Circuits and Systems and a Technical Program Committee Member for many EDA conferences. Dr.~Najafi is a Senior Member of IEEE and a Senior Member of the U.S. National Academy of Inventors (NAI). 
\end{IEEEbiography}

\begin{IEEEbiography}[{\includegraphics[width=1.09in,height=1.2in]{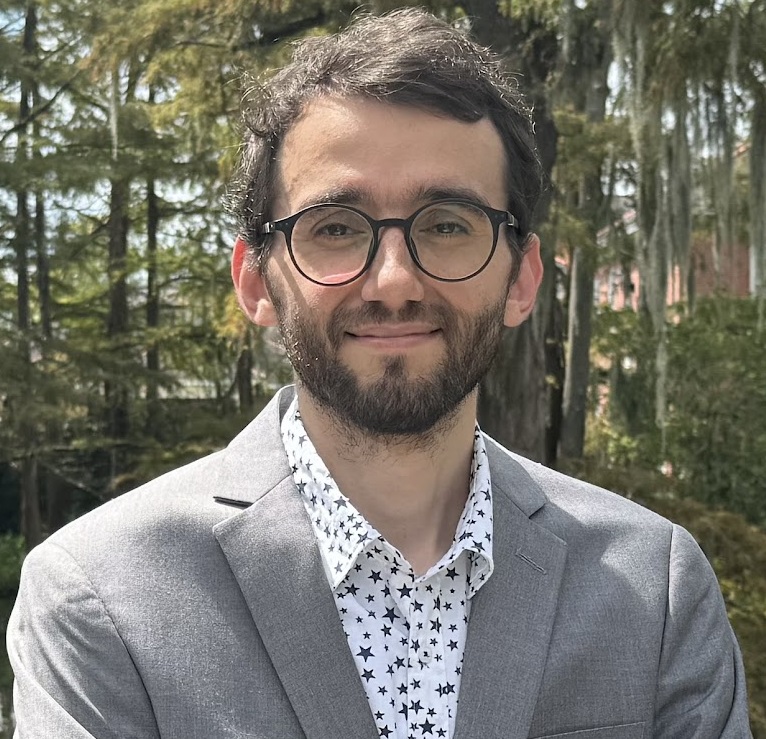}}]{Sercan Aygun} (S’09-M’22-SM'24) received a B.Sc. degree in Electrical \& Electronics Engineering and a double major in Computer Engineering from Eskisehir Osmangazi University, Turkey, in 2013. He completed his M.Sc. degree in Electronics Engineering from Istanbul Technical University in 2015 and a second M.Sc. degree in Computer Engineering from Anadolu University in 2016. Dr. Aygun received his Ph.D. in Electronics Engineering from Istanbul Technical University in 2022. %Dr. Aygun’s Ph.D. work has appeared in several Ph.D. Forums of top-tier conferences, such as DAC, DATE, ASP-DAC, and ESWEEK.
Dr. Aygun received the Best Scientific Research Award of the ACM SIGBED Student Research Competition (SRC) ESWEEK 2022, the Best Paper Award at GLSVLSI'23, and the Best Poster Awards at GLSVLSI'24 \& DCAS'26. Dr. Aygun's Ph.D. work was recognized with the Best Scientific Application Ph.D. Award by the Turkish Electronic Manufacturers Association and was also ranked first nationwide in the Science and Engineering Ph.D. Thesis Awards by the Turkish Academy of Sciences. He is currently an Assistant Professor at the School of Computing and Informatics, University of Louisiana at Lafayette, Lafayette, LA, USA. He works on tiny machine learning and emerging computing, including stochastic and hyperdimensional computing. \end{IEEEbiography}

\begin{IEEEbiography}[{\includegraphics[width=1.1in,height=1.25in,clip,keepaspectratio]{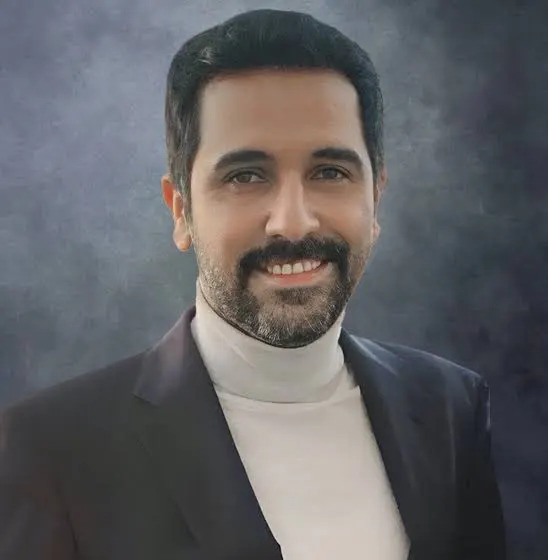}}]
{Arman Roohi} (Senior Member, IEEE) received
the Ph.D. degree in computer engineering from the
University of Central Florida, Orlando, FL, USA, in
2019. He is currently an Associate Professor with
the Department of Electrical and Computer Engineering, University of Illinois Chicago, Chicago,
IL, USA. Before joining UIC, he was an Assistant
Professor with the University of Nebraska-Lincoln
(2020–2024) and a Postdoctoral Research Fellow
with the University of Texas at Austin (2019–2020).
His research interests include the areas of crosslayer co-design for implementing complex machine learning tasks and secure
computation, including hardware security and the security of artificial intelligence, reconfigurable and adaptive computer architectures, and beyond CMOS
computing. He has completed over 50 publications on these topics, including
best paper recognition, book chapters, and STEM curricular development. He
received Ph.D. Forum at DAC 2018 Scholarship, Frank Hubbard Engineering
Endowed Scholarship in 2018, best paper recognition in IEEE TRANSACTIONS
ON EMERGING TOPICS IN COMPUTING in 2019, and paper of the month at IEEE
TRANSACTIONS ON COMPUTERS in 2017.
\end{IEEEbiography}

\begin{IEEEbiography}[{\includegraphics[width=1in,height=1.25in,clip,keepaspectratio]{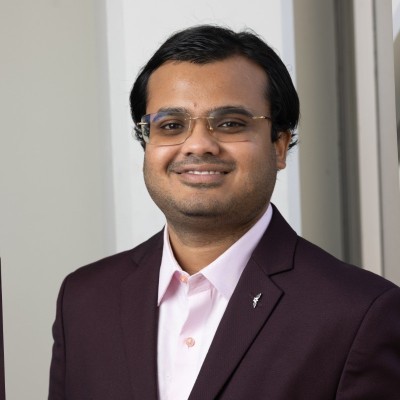}}]
{Dr. Gourav Datta} is an Assistant Professor with the ECSE Department of Case Western Reserve University (CWRU). Before that, he was an Applied Scientist at Amazon AGI, where he contributed to enhancing the video understanding capabilities of Amazon Nova, a next-generation foundation model. He obtained his PhD degree from University of Southern California in 2023. His research interests include energy and latency efficient algorithm-hardware co-design for machine learning at the edge. During his PhD tenure, he received the USC Viterbi School of Engineering’s highest achievement award, The 2024 William Ballhaus Best PhD Dissertation Award, the IEEE Graduate Fellowship on Applied Superconductivity 2022, the best Research Assistant (RA) award from USC ECE, and the Annenberg fellowship. He has published more than 60 peer-reviewed papers in top-tier venues including Nature Scientific Reports, Frontiers in Neuroscience, ICLR, ECCV, TCAS-I, DATE, WACV, among others, and received a best paper award nomination at VLSI-SoC 2022. Lastly, his research on in-sensor computing has been highlighted multiple times by Edge Impulse, a leading semiconductor IP company for TinyML, and USC Viterbi School of Engineering. 
\end{IEEEbiography}

\end{document}